\documentclass[11pt]{article}

\usepackage[letterpaper]{geometry}
\usepackage{mathtools, amsmath, amssymb}
\usepackage{tensor}
\usepackage{bbold}
\usepackage[
	  colorlinks=true,
      linkcolor=blue,  
      urlcolor=black,    
      filecolor=blue,     
      citecolor=red,
	  linktocpage=true,
      pdfstartview=FitV,
      bookmarksopen=true,
      pdftitle={A Practical Approach to the Hamilton-Jacobi Formulation of Holographic Renormalization}]{hyperref}
\usepackage{datetime}
\usepackage{authblk}

\numberwithin{equation}{section}

\def\be{\begin{equation}}
\def\ee{\end{equation}}
\def\bea{\begin{eqnarray}}
\def\eea{\end{eqnarray}}

\def\pa{\partial}   
\def\d{\delta}   
\newcommand{\eps}{\epsilon}

\newcommand{\reef}[1]{(\ref{#1})}

\title{\bf A Practical Approach to the Hamilton-Jacobi Formulation of Holographic Renormalization}
\author{\bf Henriette Elvang}
\author{\bf Marios Hadjiantonis}
\affil{Randall Laboratory of Physics, Department of Physics,\\
and Michigan Center for Theoretical Physics,\\
	   University of Michigan, Ann Arbor, MI 48109, USA}
\affil{\texttt{\href{mailto:elvang@umich.edu}{elvang@umich.edu} and \href{mailto:mhadjian@umich.edu}{mhadjian@umich.edu}}}
\date{}

\begin{document}

\begin{titlepage}

\begin{flushright}
\texttt{MCTP-16-06}
\end{flushright}

\bigskip

{\let\newpage\relax\maketitle}

\maketitle

\bigskip

\begin{abstract}
We revisit the subject of holographic renormalization for asymptotically AdS spacetimes. For many applications of holography, one has to handle the divergences associated with the on-shell gravitational action. The brute force approach uses the Fefferman-Graham (FG) expansion near the AdS boundary to identify the divergences, but subsequent reversal of the expansion is needed to construct the infinite counterterms. While in principle straightforward, the method is cumbersome and application/reversal of FG is formally unsatisfactory. Various authors have proposed an alternative method based on the Hamilton-Jacobi equation. However, this approach may appear to be abstract, difficult to implement, and in some cases limited in applicability. In this paper, we clarify the Hamilton-Jacobi formulation of holographic renormalization and present a simple algorithm for its implementation to extract cleanly the infinite counterterms. While the derivation of the method relies on the Hamiltonian formulation of general relativity, the actual application of our algorithm does not. The work applies to any $D$-dimensional holographic dual with asymptotic AdS boundary, Euclidean or Lorentzian, and arbitrary slicing.  We illustrate the method in several examples, including the FGPW model, a holographic model of 3d ABJM theory, and cases with marginal scalars such as a dilaton-axion system.

\end{abstract}

\thispagestyle{empty}

\end{titlepage}

\tableofcontents

\newpage 
\section{Introduction}

\label{sec:Intro}

In many applications of gauge-gravity duality, there is a need to regulate divergences that appear near the boundary of the bulk theory; these are simply associated with UV divergences in the dual quantum field theory.  The divergences  appear, for example, in calculations of conformal anomalies, correlation functions, and the free energy. The prescription for regulating divergences is to include suitable local counterterms. The resulting process of holographic renormalization is an old subject: it was discussed in the early days of AdS/CFT \cite{Witten:1998qj} and implemented in the classic calculations of conformal anomalies \cite{Henningson:1998gx}, the trace of the stress-tensor \cite{Balasubramanian:1999re}, and since then in countless other examples. 

We focus on bulk spacetimes that are asymptotically AdS or Euclidean AdS. This includes duals of conformal theories (CFTs) as well as holographic renormalization group flows with a UV CFT. For a given gravity dual, the local counterterms are universal and one can calculate them once and for all in any given gravitational model. We distinguish between {\em infinite counterterms} and {\em finite counterterms}. The former are unambiguous and can be determined using the bulk equations of motion. The finite counterterms, however, can typically only be fixed using further constraints, such as supersymmetry. {\em In this paper, we are concerned only with the infinite counterterms.}

There is a standard `brute force' procedure for determining the infinite counterterms \cite{Henningson:1998gx,Balasubramanian:1999re,deHaro:2000vlm,Bianchi:2001kw}. One expands the metric and fields near the AdS boundary using the Fefferman-Graham (FG) expansion \cite{FG}. Solving the equations of motion relates various coefficients in the FG expansion, but leaves unfixed the coefficients that correspond to the source and vev rates for each field. Using a suitable cutoff, the on-shell action is evaluated near the AdS boundary by plugging in the FG expansion, subject to the  equations of motion. This identifies the divergences, however, they will be expressed in terms of the free coefficients in the FG expansion. This is not sufficient, as local counterterms must be expressed directly in terms of the fields on the cutoff surface. So starting with the most divergent terms, one works systematically backwards to convert each divergence to a local field expression, thus basically reversing the FG expansion. This process identifies the field polynomials that are responsible for the divergences in the on-shell action. The counterterm action is then taken to be exactly minus those field expressions; this ensures that the renormalized action $S_\text{bulk} + S_\text{ct}$ is finite. 
(This still leaves the possibility of ambiguities from finite counterterms; we will discuss this briefly in the Discussion section.)

While straightforward for many simple models with just one or two scalar fields, the brute force approach outlined above becomes increasingly tedious for models with multiple fields. Moreover, it is fundamentally unsatisfying that one first abandons the field expressions in favor of Fefferman-Graham only to reverse back to fields after identifying the infinite terms. For this reason, another approach, based on the Hamiltonian formalism for gravity and the Hamilton-Jacobi equation,  has been proposed for   holographic renormalization. 

Early in the studies of holographic renormalization group flows, de Boer, Verlinde, and Verlinde \cite{deBoer:1999tgo} proposed to use the Hamilton-Jacobi equation to derive first-order equations for the supergravity model and they related it to the Callan-Symanzik equation. (See also \cite{Verlinde:1999xm,Verlinde:1999mg} and the lectures \cite{deBoer:2000cz}.) The  specific application of the Hamilton-Jacobi equation to determine infinite counterterms  was  studied by Kalkkinen, Martelli, and Mueck in \cite{Kalkkinen:2001vg,Martelli:2002sp} and subsequently by Papadimitriou and Skenderis in \cite{Papadimitriou:2004ap} (see also \cite{Papadimitriou:2004rz,Papadimitriou:2010as,Papadimitriou:2011qb}). 

One limitation of the method as formulated   
in \cite{Papadimitriou:2004ap} is that the dilatation operator is used to organize the calculation. This requires that the fields are eigenfunctions of the dilatation operator, but that makes it more challenging to handle scalars dual to operators with scaling dimension $\Delta= d/2$, because of their leading $\log$-falloff.\footnote{One can work around this, see for example \cite{Papadimitriou:2004rz}. The issue is also addressed in \cite{Papadimitriou:2011qb}.} This is not an exotic case, but a very common one; for example, in a $d=4$ field theory, a scalar mass term is a relevant operator of dimension $\Delta=2$.  
Another challenge is that, as presented in \cite{Papadimitriou:2004ap}, the Hamilton-Jacobi method looks rather difficult to carry out in practice. 

The goal of this paper is to straighten out and simplify the Hamilton-Jacobi approach for holographic renormalization.  We will show that the application of the Hamilton-Jacobi equation  
\be \label{HJintro}
   \frac{\partial S_\text{on-shell}}{\partial r }  + H = 0
\ee
(with the radial coordinate $r$ playing the role of the usual time-coordinate), 
can be implemented via an algorithm that significantly simplifies the process of computing the infinite counterterms. To avoid the issue of the dilatation operator and have an approach that applies more generally, we organize the calculation in terms of a derivative expansion (or inverse metric expansion), 
as also suggested in for example \cite{deBoer:1999tgo,Martelli:2002sp,Papadimitriou:2011qb}.

We will be working with bulk actions of the form
\begin{equation}
S = - \frac{1}{2	 \kappa^2} \int_M d^{d + 1}x \sqrt{g}\, \Big ( \mathcal{R}[g] - g^{\mu \nu} G_{IJ} \partial_\mu \Phi^I \partial_\nu \Phi^J - V ( \Phi ) \Big ) \,,
\end{equation}
where we allow for a general metric $G_{IJ}=G_{IJ}(\Phi)$ on the scalar manifold. We  consider domain wall solutions with arbitrary slicing and assume that the asymptotic UV structure of the metric is AdS (or Euclidean AdS). For such a system, we formulate the Hamilton-Jacobi problem for the on-shell action $S_\text{on-shell}$; \reef{HJintro} 
 is basically a partial differential equation for $S_\text{on-shell}$ and once derived, one no longer has to think about the Hamiltonian formulation of general relativity. Instead, one systematically solves the Hamilton-Jacobi differential equation for $S_\text{on-shell}$ by writing a suitable Ansatz for its divergent terms and then solving for the coefficients in this Ansatz.  The key point here is that scalars dual to relevant operators in the field theory go to zero at the boundary. Therefore there can only be limited powers of each field in the infinite counterterms, and that makes the Ansatz finite. 

Our method departs from previous approaches as follows.\footnote{However, see  \cite{Larsen:2003pf} for a similar approach in dS space.} We consider $S_\text{on-shell}$ as the action on the cut-off boundary; this breaks the general diffeomorphism invariance in the radial direction and therefore we must take seriously the explicit dependence on the radial coordinate in $S_\text{on-shell}$. Thus, the $r$  partial-derivative in \reef{HJintro}  plays a central role in our method. In fact, the coefficients in our Ansatz will be allowed to have explicit $r$-dependence, and the Hamilton-Jacobi equation then yields differential equations for these coefficients that we can solve  unambiguously in the near boundary limit.

We illustrate the use of the method in several contexts. To start out, we reproduce the purely gravitational counterterms \cite{deHaro:2000vlm} in $d$-dimensions. To show how the method works for a case with $d$ odd, we reproduce the infinite counterterms of the $d=3$ ABJM dual model of \cite{Freedman:2013ryh}.  We then turn to the example of the $d=4$ FGPW model \cite{Freedman:1999gp} whose two scalars have $\Delta =2$ and $\Delta=3$.

In the presence of a marginal scalar, more care must be taken. A marginal scalar generically goes to a finite value at the boundary and therefore the associated counterterms do not enjoy the same suppression as the scalars dual to relevant operators. We handle  this by allowing the coefficients of our Ansatz for $S_\text{on-shell}$ to be functions of the marginal scalar. We have applied this method successfully to calculate the counterterms for a ten-scalar model dual to (a limit of) $\mathcal{N}=1^*$ theory on $S^4$ \cite{Bobev:2016nua}; this indeed served as a motivation for us to revisit the subject of holographic renormalization. However, for the purpose of presentation here, we restrict ourselves to simply show how our method reproduce the infinite counterterms for the dilaton-axion system  in \cite{Papadimitriou:2011qb}.

The paper is organized as follows. In Section \ref{sec:generalHJ}, we present the Hamilton-Jacobi equation for the bulk and describe our algorithm for determining the infinite counterterms. Section \ref{sec:PureGravity} implements the method for pure gravity in $d$ dimensions. The examples of the ABJM model and FGPW can be found in Sections \ref{sec:ABJMmodel} and \ref{sec:FGPW}; these give very concrete illustrations of how we implement the algorithm. The more advanced case of marginal scalars is treated in Section 
\ref{sec:dilaton}. The three appendices contain various technical details. Appendix \ref{app:formulas} is a short list of useful identities for the metric variations of gravitational curvatures. Appendix \ref{app:SixDerivatives} gives details of the calculation of the gravitational six-derivative terms needed for counterterms in $d=6$. Finally, Appendix \ref{app:1ptFunctions} offers explicit calculation of the one-point functions in FGPW to illustrate that the one-point functions determined from the renormalized action with our infinite counterterms are indeed all finite.

\section{Hamiltonian Approach to Holographic Renormalization}
\label{sec:generalHJ}

We start with a brief description of the essential parts of the Hamiltonian formulation needed for holographic renormalization. We then formulate the problem of determining the on-shell action in terms of the Hamilton-Jacobi equation and we present our algorithm for calculating the divergent part of the on-shell action.

\subsection{Hamiltonian formalism of gravity}

We consider a general form of the bulk gravitational action:
\begin{equation}
S = - \frac{1}{2	 \kappa^2} \int_M d^{d + 1}x\ \sqrt{g} \left ( \mathcal{R}[g] - g^{\mu \nu} G_{IJ} \partial_\mu \Phi^I \partial_\nu \Phi^J - V ( \Phi ) \right ) - \frac{1}{\kappa^2} \int_{\partial M} d^d x\ \sqrt{\gamma} K\,.
\label{Sbulk}
\end{equation}
The last term in \reef{Sbulk} is the Gibbons-Hawking boundary term which ensures that the variational problem is well-defined. In this term, $\gamma_{ij}$ is the induced metric on the boundary and $K$ is its extrinsic curvature.

We choose a gauge for the bulk metric $g_{\mu\nu}$ such that the line element takes the form
\begin{equation}
\label{lineelement}
ds^2 = dr^2 + \gamma_{ij} ( r, x ) dx^i dx^j\,,
\end{equation}
where latin indices $i, j,\dots$ are in the range $i, j = 1, 2, \ldots, d$ and will denote boundary coordinates.

This allows us to decompose the Ricci scalar in the action to get
\begin{equation}
S = - \frac{1}{2 \kappa^2} \int_M d^dx\ dr \sqrt{\gamma} \Big ( R[\gamma] + K^2 - K_{ij} K^{ij} 
         - G_{IJ} \dot{\Phi}^I \dot{\Phi}^J - \gamma^{ij} G_{IJ} \partial_i \Phi^I \partial_j \Phi^J - V ( \Phi ) 
\Big )\,,
\label{SbulkK}
\end{equation}
where the extrinsic curvatures are 
\begin{equation}
\label{Ks1}
\tensor{K}{^i_j} = \frac{1}{2} \gamma^{ik} \dot{\gamma}_{kj} \qquad \textrm{and} \qquad K = \frac{1}{2} \gamma^{ij} \dot{\gamma}_{ij} \,.
\end{equation}
The dots denote derivatives with respect to $r$.
The boundary Gibbons-Hawking term does not appear in the expression \reef{SbulkK}, since it has been canceled by boundary terms that occur from partial integration of second derivative terms in the expansion of $\mathcal{R}[g]$.

In the Hamiltonian formulation of holographic renormalization, the radial coordinate $r$  plays the role of the time coordinate. Therefore, the conjugate momenta  to the fields are given by
\begin{equation}
  \pi^{ij} = \frac{\delta S}{\delta \dot{\gamma}_{ij}} = \frac{1}{2 \kappa^2} \sqrt{\gamma} \left    
  ( K^{ij} - K \gamma^{ij} \right ) 
  \qquad \textrm{and} \qquad 
  \pi_I = \frac{\delta S}{\delta \dot{\Phi}^I} 
  = \frac{1}{\kappa^2} \sqrt{\gamma} G_{IJ} \dot{\Phi}^J\,,
\label{eq:HamiltonMomenta}
\end{equation}
and the Hamiltonian is
\begin{equation}
\begin{aligned}
H & = \int_{\partial M} d^dx \left ( \pi^{ij} \dot{\gamma}_{ij} + \pi_I \dot{\Phi}^I - \mathcal{L} \right ) \\
& = \frac{1}{2 \kappa^2} \int_{\partial M} d^dx \sqrt{\gamma} \left ( R[\gamma] - K^2 + K_{ij} K^{ij} + G^{IJ} p_I p_J - \gamma^{ij} G_{IJ} \partial_i \Phi^I \partial_j \Phi^J - V ( \Phi ) \right )\,,
\end{aligned}
\end{equation}
where, for simplicity, we have introduced $p_I \equiv \frac{\kappa^2}{\sqrt{\gamma}}\pi_I$.

\subsection{Hamilton-Jacobi formulation}
The Hamilton-Jacobi formulation is well-known in classical mechanics \cite{Goldstein}. With  the radial coordinate $r$ playing the role of time, the  Hamilton-Jacobi equation takes the form
\begin{equation}
H + \frac{\partial S_\text{on-shell}}{\partial r} = 0\,.
\label{eq:HJEquation}
\end{equation}
Just as in classical mechanics, it is key to emphasize that in the Hamilton-Jacobian formalism, the Hamiltonian is a functional of canonical momenta {\em defined} by
\be
  \label{conjmomS}
  \pi^{ij} = \frac{\delta S_\text{on-shell}}{\delta \gamma_{ij}}  
  ~~~~\text{and}~~~~
  p_I 
  =
  \frac{\kappa^2}{\sqrt{\gamma}}
  \pi_I 
 = 
   \frac{\kappa^2}{\sqrt{\gamma}}
  \frac{\delta S_\text{on-shell}}{\delta \Phi^I}    \,,
\ee
as opposed to the canonical definitions \reef{eq:HamiltonMomenta}. When the momenta are defined via equation~(\ref{eq:HamiltonMomenta}) with the extrinsic curvature given by \reef{Ks1}, the Hamiltonian constraint of Einstein's equation is simply $H=0$. If this were used with the Hamilton-Jacobi equation \reef{eq:HJEquation}, it would imply that the action has no explicit $r$-dependence; this is of course true for the diffeomorphism-invariant  gravitational bulk action whose metric equations-of-motion imply the Hamiltonian constraint. However, it is not true for the {\em on-shell action}, which is an action on the cut-off boundary. It has explicit $r$-dependence, as we shall see, and to determine it via the Hamilton-Jacobi equation we must use the definitions \reef{conjmomS}. With \reef{conjmomS}, the Hamilton-Jacobi equation \reef{eq:HJEquation} should be thought of as a first-order partial differential equation for $S_\text{on-shell}$ with respect to the fields, the metric, and $r$.

A practical approach is to use an Ansatz for the on-shell action: below we will be more explicit about how we choose an appropriate Ansatz, but for now we will develop the general formalism further. Let us write the Ansatz as
\begin{equation}
\label{Sansatz}
S_\text{on-shell} = \frac{1}{\kappa^2} \int_{\partial M_\eps} d^d x \sqrt{\gamma}\, U ( \gamma, \Phi, r ) \,.
\end{equation}
The function $U$ is a function of the induced (inverse) metric $\gamma^{ij}$ on the boundary and the scalar fields $\Phi^I$, and it has also explicit dependence on $r$.
The cutoff surface $\partial M_\eps$ becomes the boundary of the spacetime when $\eps \to 0$.

Using the above Ansatz, the Hamilton-Jacobi equation takes the form
\begin{equation}
R[\gamma] +  K_{ij}K^{ij}- K^2  + G^{IJ} p_I p_J - \gamma^{ij} G_{IJ} \partial_i \Phi^I \partial_j \Phi^J - V ( \Phi ) + 2 \frac{\partial U}{\partial r} \,=\, 0\,.
\label{eq:HJEquationFinal}
\end{equation}
We emphasize that this equation is to be understood as an integral equation, i.e.~it holds up to total derivatives and we can manipulate it using partial integration in the boundary coordinates. 

As discussed above, the conjugate momenta in \reef{eq:HJEquationFinal} will be given by derivatives of $U$. For the scalar field conjugates, this straightforwardly gives
\be
  p_I 
  = \frac{\,\kappa^2}{\sqrt{\gamma}} \frac{\delta S_\text{on-shell}}{\delta \Phi^I}  
  ~~\Rightarrow ~~ 
  \boxed{ ~p_I = \frac{\delta U}{\delta \Phi^I}\,. ~}\
  \label{pI}
\ee
The conjugate momentum of the metric enters  \reef{eq:HJEquationFinal} via the extrinsic curvatures, since $K^{ij} = \frac{2\kappa^2}{\sqrt{\gamma}} \big( \pi^{ij} - \frac{1}{d-1} \gamma^{ij} \pi^{kl} \gamma_{kl}\big)$, as follows from \reef{eq:HamiltonMomenta}. 
Now in the context of the Hamilton-Jacobi formalism, the extrinsic curvatures $K^{ij}$ in \reef{eq:HJEquationFinal} must then be expressed in terms of $\pi^{ij}$ as given by \reef{conjmomS}. This gives 
\begin{equation}
\tensor{K}{^i_j} = - 2 \gamma^{ik} \frac{\delta U}{\delta \gamma^{kj}} - \frac{1}{d - 1} \left ( U - 2 \gamma^{mn} \frac{\delta U}{\delta \gamma^{mn}} \right ) \tensor{\delta}{^i_j}\,,
\label{eq:JacobiGammaMomentum}
\end{equation}
where we have used $\gamma_{ij} \gamma^{jk} = \tensor{\delta}{_i^k} ~\implies~(\delta \gamma_{ij} )\gamma^{jk} = -\gamma_{ij} (\delta \gamma^{jk})$ to express $\tensor{K}{^i_j}$ in terms of derivatives with respect to the {\em inverse} metric rather than the metric; this will be useful later. 

It is convenient to define 
\be
   \boxed{
   ~~Y_{ij} = \frac{\delta U}{\delta \gamma^{ij}} 
   ~~~~\text{and}~~~~ 
   Y = \gamma^{ij} Y_{ij}\, .~
   }
\ee
One then finds from \reef{eq:JacobiGammaMomentum} that the dependence on extrinsic curvatures in the Hamilton-Jacobi equation \reef{eq:HJEquationFinal} is given in terms of $U$ as 
\be
  \boxed{
  ~~\mathcal{K} ~\equiv~ 
 ~K_{ij}K^{ij} - K^2 ~=~ 4 Y_{ij} Y^{ij} - \frac{1}{d - 1} (U - 2 Y)^2 - U^2\,.~
  }
  \label{KKinHJ}
\ee

To summarize, our strategy for computing the on-shell action $S_\text{on-shell}$ is to use the Ansatz \reef{Sansatz} and  solve the Hamilton-Jacobi equation 
\begin{equation}
\boxed{
~~R[\gamma] +  \mathcal{K}  + G^{IJ} p_I p_J - \gamma^{ij} G_{IJ} \partial_i \Phi^I \partial_j \Phi^J - V ( \Phi ) + 2 \frac{\partial U}{\partial r} = 0\,.~
\label{eq:HJEquationFinal2}
}
\end{equation}
with conjugate momenta given by \reef{pI} and $\mathcal{K}$ defined in \reef{KKinHJ}.
We remind the reader that equation \reef{eq:HJEquationFinal2} has to hold only as an integral equation, so we are free to manipulate it using partial integration. 
While this was derived using the Hamiltonian formalism of gravity, we no longer need to think of the problem that way. Rather, we now have differential equation \reef{eq:HJEquationFinal2} for the on-shell action $S_\text{on-shell}$. Next, we explain how to solve it systematically.

\subsection{Algorithm to determine the divergent part of the on-shell action}
\label{s:algo}

Let us next address how we propose to use the Hamilton-Jacobi formulation to determine the divergent part of the on-shell action and thereby the counterterms needed for a finite result. We outline here the general approach, however the method is much better illustrated by concrete examples; these follow in the next sections. 

We assume that asymptotically the bulk metric approaches AdS space: in terms of the choice of coordinates  \reef{lineelement}, 
$ds^2 = dr^2 + \gamma_{ij} ( r, x ) dx^i dx^j$, this means that 
\be
  \label{asympmetric}
  \gamma_{ij} \to e^{2r/L}\, \gamma_{(0)ij} + \ldots
  ~~~~\text{as}~~~~
 r \to \infty \,,
\ee
where $L$ is the AdS radius. 
The boundary metric $\gamma_{(0)ij}$ can be Lorentzian or Euclidean, it can be flat or curved. For example, recent applications of holography considered the dual field theory on $d$-dimensional compact Euclidean spaces, such as spheres. In the following,  $\gamma_{(0)ij}$ will be general. 

The asymptotic behavior \reef{asympmetric}, gives $\sqrt{\gamma} \sim e^{d r} \sqrt{\gamma_{(0)}}$. We are focusing only on the divergent parts of the on-shell action, so we need terms in $U$ only up to orders $e^{-d r}$ (possibly including also terms polynomial in $r$). Since the inverse metric  $\gamma^{ij}$ scales as 
$e^{-2r}$, we can ignore any terms with more than $\left \lfloor \frac{d}{2} \right \rfloor$ inverse metrics.  Any (boundary) derivatives that appear in terms in $U$ must necessarily be contracted pairwise by inverse metrics $\gamma^{ij}$, so we do not consider terms with more than $d$-derivatives. All in all, this makes it natural to organize  the Ansatz for $U$ in a derivative expansion:
\begin{equation}
U = U_{(0)} + U_{(2)} + \ldots + U_{\left ( 2 \left \lfloor \frac{d}{2} \right \rfloor \right )} \,,
\end{equation}
where the subscript represents the number of derivatives in each term.
Curvature terms such as the boundary Ricci scalar, Ricci tensor, and Riemann tensor are each order 2 (i.e. they have two derivatives). Previous work, for example \cite{deBoer:1999tgo} and \cite{Papadimitriou:2011qb}, have also organized the on-shell action as a derivative expansion.

For the 0th order in the derivative expansion, we have $Y_{(0)ij} = \frac{\delta U_{(0)}}{\delta \gamma^{ij}}  = 0$, so \reef{KKinHJ} simply gives 
\be
 \label{K0}
 \mathcal{K}_{(0)} ~=~  - \frac{d}{d - 1} U_{(0)}^2 \,.
\ee 
Thus at 0th order, the Hamilton-Jacobian equation \reef{eq:HJEquationFinal} becomes
\be
  \label{fakeSP}
  V ( \Phi )= 
 G^{IJ} \frac{\delta U_{(0)}}{\delta \Phi^I} \frac{\delta U_{(0)}}{\delta \Phi^J} 
- \frac{d}{d - 1} U_{(0)}^2   
+ 2 \frac{\partial U_{(0)}}{\partial r} \,.
\ee
Without the last $r$-derivative term, we see that $U_{(0)}$ is essentially like a (fake) superpotential for the scalar potential $V$; this was also noted  \cite{deBoer:1999tgo} (see also \cite{Mueck:2001cy,Papadimitriou:2011qb}). In general, it is not easy to solve for a superpotential for a given $V$; however, we will not need to since our focus is on the generic asymptotically divergent terms only. As noted in the discussion below \reef{conjmomS} the presence of the explicit $r$-derivative term in the Hamilton-Jacobi equation, and hence in \reef{fakeSP}, is crucial --- this point does not seem to have been appreciated in previous discussions of the method. 

Let us for later convenience also record the results for $\mathcal{K}$ at two- and four-derivative order:
\be
 \begin{split}
   \label{K2K4}
   \mathcal{K}_{(2)} &~=~ 
   - \frac{2}{d-1} U_{(0)} \big[U_{(2)} - 2 Y_{(2)}\big] - 2 U_{(0)} U_{(2)} \,,
   \\
   \mathcal{K}_{(4)} &~=~  
   4 Y_{(2)ij} Y^{ij}_{(2)}
   - \frac{1}{d-1} \big[U_{(2)} - 2 Y_{(2)}\big]^2 
   -\frac{2}{d-1}U_{(0)} \big[U_{(4)} - 2 Y_{(4)}\big]
   -  U_{(2)}^2
   - 2 U_{(0)} U_{(4)} \,,
 \end{split}
\ee 
where $Y_{(k)ij} = \frac{\delta U_{(k)}}{\delta \gamma^{ij}}$.

\vspace{5mm}
\noindent {\em 
We propose the following algorithm to determine the infinite terms in the on-shell action:}

\paragraph{Step 1: Ansatz for $U_{(2n)}$.} For each $U_{(2n)}$, we write a systematic Ansatz that includes all 
potentially divergent terms of this order with undetermined coefficients,\footnote{Terms are considered equivalent if related by partial integration.} for example
\be 
\label{U0example}
U_{(0)} = A_0 + A_1 \phi + A_3 \phi^2 + \ldots
~~~~\text{and}~~~~
U_{(2)} = B_0 R + B_1 R \phi + B_2 \phi \Box \phi + \dots 
\ee
where the coefficients $A_i$ and $B_i$ can have explicit dependence on $r$.  The Hamilton-Jacobi equations will therefore give us differential equations of these coefficients which we solve asymptotically, keeping only terms that give divergent contributions to the on-shell action. 

Recall that the asymptotic behavior of a scalar with bulk mass $m_I^2$ is $\Phi^I \to  \Phi_{(0)}^I e^{- (d-\Delta_I) r / L}$, where $m_I^2 L^2 = \Delta_I ( \Delta_I - d )$. The two  solutions for $\Delta_I$ correspond to the source and vev-rate falloffs. When a scalar approaches zero at the boundary, as is the case in many applications, we can immediately read off how many powers of the scalar can possibly appear in $U_{(2n)}$; the number of possible terms is finite and limited by the fact that we are only interested in the divergent terms.\footnote{We will also discuss cases with a marginal scalar $m_I^2 = 0$, for which there is no suppression near the boundary and generically the scalar goes to a non-zero constant. For such cases, we allow the coefficients $A_i$ in our Ansatz to be functions of the marginal scalar. An example  is presented in Section \ref{sec:dilaton}.}  For example, if $\phi$ is a scalar with dimension $\Delta_\phi=3$ in $d=4$, then $\phi \sim e^{-r}$, and we have to include powers up to $\phi^4$ in $U_{(0)}$ and $\phi \Box \phi$ can appear in $U_{(2)}$. (Note: such terms with $e^{-dr}$ falloff will be finite unless the $r$-dependence in the coefficient makes it divergent.) On the other hand, if $\phi$ in \reef{U0example} is a $\Delta_\phi=2$ scalar in $d=4$, there can at most be quadratic powers of $\phi$ in $U_{(0)}$ and the term $\phi \Box \phi$ is not divergent, so it is not included in the Ansatz for $U_{(2)}$.

One can  impose symmetries of the theory in order to further simplify the Ansatz for $U_{(2n)}$. If, for example, the bulk action has a symmetry $\phi \to -\phi$, we can drop any terms odd under this symmetry in the Ansatz.

\paragraph{Step 2: Conjugate momenta.} Next, using the leading asymptotic behaviors of the fields, we determine the leading asymptotics of the conjugate momenta.
Using this together with $p_I = \frac{\delta U}{\delta \Phi^I}$ fixes some of the coefficients in $U_{(0)}$ quite easily.

\paragraph{Step 3: Solving the Hamilton-Jacobi equation.} We plug the Ansatz for $U_{(2n)}$  into the Hamilton-Jacobi equation and we solve it order by order by demanding that the coefficients of the different field monomials vanish independently. When necessary, use partial integration to eliminate potentially non-independent terms that appeared by varying $U$. We start with $U_{(0)}$, then use those results to determine $U_{(2)}$, then $U_{(4)}$ etc. 

\paragraph{Step 4: Counterterm action.} Once the divergent terms in $S_\text{on-shell}$ have been determined, the counterterm action is simply 
\be 
S_\text{ct} = -S_\text{on-shell}\big|_\text{div}\,.
\ee 
This is added to the bulk action to get the regularized action $S_\text{reg} = S_\text{bulk} + S_\text{GH} + S_\text{ct}$ from which correlation functions can be computed and by construction are guaranteed to be finite. In many cases, counterterm actions are presented in term of the Fefferman-Graham radial coordinate $\rho$ related to $r$ via $\rho = e^{-2r/L}$, so that the line element is 
\begin{equation}
  ds^2 = L^2 \frac{d \rho^2}{4 \rho^2} + \gamma_{ij} \, dx^i dx^j \,.
\end{equation}
We determine the divergent terms in the on-shell action using the $r$-coordinate, but convert to $\rho$-coordinates for the final presentation of our counterterm actions. In terms of the $\rho$-coordinate, the cutoff surface $\partial M_\eps$, introduced in \reef{Sansatz}, is then located at $\rho = \eps$.

\vspace{5mm}
In the following sections, we demonstrate the procedure explicitly in a set of representative explicit  examples. We start with pure gravity in $d$-dimensions with $d=2,3,4,5,6$, then move on to a $d=3$ ABJM dual model and the $d=4$ two-scalar model known as FGPW. 
Finally, we illustrate how our method works with marginal scalars (dilaton + axion in $d=4$).

\section{Pure Gravity}
\label{sec:PureGravity}

The simplest model one can consider is pure AdS gravity with no matter content in $D = d + 1$ dimensions.
Counterterms obtained by renormalizing this model will be present in every other model and it is therefore useful to  deal with them once and for all.
The action we consider is given by \reef{Sbulk} with no scalar fields and constant scalar potential 
\begin{equation}
V =  -\frac{d ( d - 1 )}{L^2} \,.
\end{equation}
The Hamilton-Jacobi equation~(\ref{eq:HJEquationFinal2}) simplifies to 
\begin{equation}
R [ \gamma ] +\mathcal{K} + \frac{d ( d - 1 )}{L^2} + 2 \frac{\partial U}{\partial r} = 0\,,
\label{eq:HJEquationPureGravity}
\end{equation}
with $\mathcal{K}$ given by \reef{KKinHJ}. Let us now apply the algorithmic procedure described in the previous section in order to determine the necessary counterterms for this class of theories.

\paragraph{Step 1:}
Since there are no scalars, the general Ansatz for each order of the expansion of $U$ is\begin{equation}
U_{(0)} = A ( r )\,,~~~~~
U_{(2)} = B ( r ) R\,,~~~~~
U_{(4)} = C_1 ( r ) R_{ij} R^{ij} + C_2 ( r ) R^2\,,~~~~
\end{equation}
where the four-derivative terms are only needed for  $d \ge 4$.\footnote{In $U_{(4)}$, one could also have included a term with the square of the Riemann tensor. However, it is not hard to see that its coefficient will be set to zero in the HJ equation.} 
We are not including terms like $\square R$  since it is a total derivative and it will not contribute in the on-shell action.
For $d \ge 6$, we need 
\begin{equation}
U_{(6)} = D_1 R^3 + D_2 R R_{ij} R^{ij} + D_3 \tensor{R}{_i^j} \tensor{R}{_j^k} \tensor{R}{_k^i} + D_4 R^{ij} R^{kl} R_{ikjl} + D_5 R \square R + D_6 R_{ij} \square R^{ij} \,.
\label{eq:SixDerivativeAnsatz}
\end{equation}
This is not a complete list of independent six-derivative terms, but it turns out to be a sufficient list. 

It is important that all the coefficients in the above expressions for $U$ depend on the radial coordinate $r$, as this will capture the explicit $r$-dependence of the on-shell action.

\paragraph{Step 2:}
This step is irrelevant for the pure gravity case since there are no matter fields.

\paragraph{Step 3:}
We now solve Hamilton-Jacobi equation \reef{eq:HJEquationPureGravity} order by order to determine the unknown coefficients $A$, $B$, $C_{1,2}$ and $D_i$.

At zero-derivatives, \reef{eq:HJEquationPureGravity} with $\mathcal{K}_{(0)}$ given by \reef{K0} gives 
\be
2\dot{A} - \frac{d}{d - 1} A^2 + \frac{d ( d - 1 )}{L^2} = 0\,,
\ee
where the dot denotes differentiation with respect to $r$.
For large $r$, the solution to the  differential equation is 
\be
A ( r ) = 
 - \frac{d - 1}{L} + \mathcal{O} \big ( e^{- d r / L} \big )\,.
\ee
The subleading terms in the large-$r$ expansion of $A$ give only finite contribution to the on-shell action and we can drop it to simply have
\begin{equation}
\label{pgU0}
U_{(0)} = - \frac{d - 1}{L}\,.
\end{equation}
This captures the leading divergence associated with the cosmological constant. 

At two-derivative order, the HJ equation~(\ref{eq:HJEquationPureGravity}) with \reef{K2K4} gives
\begin{equation}
R - \frac{2}{d - 1} U_{(0)} \left ( U_{(2)} - 2 Y_{(2)} \right ) - 2 U_{(0)} U_{(2)} + 2 \frac{\partial U_{(2)}}{\partial r} = 0\,.
\end{equation}
The inverse-metric variation of $U_{(2)}$ simply gives $Y_{(2)ij} =\frac{\delta U_{(2)}}{\delta \gamma^{ij}} = B R_{ij}$, so $Y_{(2)} = B R$.
With the solution for $U_{(0)}$ in \reef{pgU0}, we obtain the following differential equation for $B$:
\be
2\dot{B} + 2\frac{d - 2}{L} B + 1 = 0\,.
\ee 
The differential equation for $B$ has solution
\be
\label{Bpuregrav}
B ( r ) = \left \{ \begin{array}{ll}
- \frac{r}{2} + \mathcal{O} ( 1 ) & \textrm{ for } d = 2 \\[0.3em]
- \frac{L}{2 ( d - 2 )} + \mathcal{O} \left ( e^{- ( d - 2 ) r / L} \right ) & \textrm{ for } d > 2
\end{array} \right .
\ee
In both cases, the subleading terms are not important since they give finite contributions to the on-shell action.
The result is therefore
\begin{equation}
U_{(2)} = \left \{ \begin{array}{ll}
- \frac{r}{2} R & \textrm{ for } d = 2 \\[0.3em]
- \frac{L}{2 ( d - 2 )} R & \textrm{ for } d > 2
\end{array} \right .
\end{equation}
The linear $r$ behavior in the $d=2$ case is our first illustration of the explicit $r$-dependence in the on-shell action and the importance of keeping the $\frac{\partial S_\text{on-shell}}{\partial r}$-term in the Hamilton-Jacobi equation. 

For the four-derivative terms, we calculate the inverse-metric variation of $U_4$ using the formulas in Appendix \ref{app:formulas}. In particular, we find $Y_{(4)} = 2 C_1 R_{ij} R^{ij}+ 2 C_2 R^2$ (up to total derivatives that can be dropped). Using this together with the results for $Y_{(2)}$ above, we can calculate $\mathcal{K}_{(4)}$ given in  \reef{K2K4}. At 4th order, the HJ equation \reef{eq:HJEquationPureGravity} is simply 
$\mathcal{K}_{(4)} + 2 \frac{\partial U_{(4)}}{\partial r} = 0$ and collecting terms gives
\begin{multline}
\left [ 2 \dot{C}_1 + \frac{2 ( d - 4 )}{L} C_1 + \left ( \frac{L}{d - 2} \right )^2  \right ] R_{ij} R^{ij} + \left [ 2 \dot{C_2} + \frac{2 ( d - 4 )}{L} C_2 - \frac{d L^2}{4 ( d - 1 ) ( d - 2 )^2} \right ] R^2 = 0\,. \nonumber
\end{multline}
Demanding the coefficients of the $R_{ij} R^{ij}$ and $R^2$ terms to vanish independently results in two differential equation for the coefficients $C_1$ and $C_2$, which have solutions
\be
C_1 = 
\left \{ \begin{array}{ll}
- \frac{L^2 r}{8} + \mathcal{O} ( 1 ) & \textrm{ for } d = 4 \\[0.3em]
- \frac{L^3}{2 ( d - 2 )^2 ( d - 4 )} + \mathcal{O} \left ( e^{- ( d - 4 ) r / L} \right ) & \textrm{ for } d > 4
\end{array} \right .
\ee
\be
C_2 = \left \{ \begin{array}{ll}
\frac{L^2 r}{24} + \mathcal{O} ( 1 ) & \textrm{ for } d = 4 \\[0.3em]
\frac{d L^3}{8 ( d - 1 ) ( d - 2 )^2 ( d - 4 )} + \mathcal{O} \left ( e^{- ( d - 4 ) r / L} \right ) & \textrm{ for } d > 4
\end{array} \right .
\ee
Again, the subleading terms can be dropped because they give only finite contributions to the on-shell action. Thus, the result for $U_{(4)}$ is 
\begin{equation}
U_{(4)} = \left \{ \begin{array}{ll}
- \frac{L^2 r}{8} \left ( R_{ij} R^{ij} - \frac{1}{3} R^2 \right ) & \textrm{ for } d = 4 \\[0.3em]
- \frac{L^3}{2 ( d - 2 )^2 ( d - 4 )} \left ( R_{ij} R^{ij} - \frac{d}{4 ( d - 1)} R^2 \right ) & \textrm{ for } d > 4
\end{array} \right .
\end{equation}
\paragraph{Step 4:}
We now have all information needed to write the counterterm action.
\be
  S_\text{ct} = - \frac{1}{\kappa^2} \int_{\partial M_\eps} d^d x \sqrt{\gamma}\, U 
   = - \frac{1}{\kappa^2} \int_{\partial M_\eps} d^d x \sqrt{\gamma}\, 
   \Big[U_{(0)} + U_{(2)} + \ldots + U_{\left ( 2 \left \lfloor \frac{d}{2} \right \rfloor \right )}\Big]\,.
\ee
Summarizing the above results,  the purely gravitational counterterms are
\be
  \boxed{
  \begin{array}{lll}
    d=2\!:~~~&\displaystyle
    S_\text{ct} 
  = \frac{1}{\kappa^2} \int_{\partial M_\eps} d^d x  \sqrt{\gamma}\, \bigg[ 
      \frac{1}{L} 
      - \log\rho ~\frac{L}{4}R
      \bigg]\,,
      \\[5mm]
    d=3\!:~~~&\displaystyle
    S_\text{ct} 
  = \frac{1}{\kappa^2} \int_{\partial M_\eps} d^d x  \sqrt{\gamma}\, \bigg[ 
      \frac{2}{L} 
      +\frac{L}{2}R
      \bigg]\,,
      \\[5mm]
    d=4\!:~~~&\displaystyle
    S_\text{ct} 
  = \frac{1}{\kappa^2} \int_{\partial M_\eps} d^d x  \sqrt{\gamma}\, \bigg[ 
      \frac{3}{L} 
      + \frac{L}{4} R 
      - \log\rho \,\,\frac{L^3}{16} \bigg( R_{ij} R^{ij} - \frac{1}{3} R^2 \bigg)
      \bigg]\,,
      \\[5mm]
    d=5\!:~~~&\displaystyle
    S_\text{ct} 
  = \frac{1}{\kappa^2} \int_{\partial M_\eps} d^d x  \sqrt{\gamma}\, \bigg[ 
      \frac{4}{L} 
      + \frac{L}{6} R 
      +\frac{L^3}{18} \bigg( R_{ij} R^{ij} - \frac{5}{16} R^2 \bigg)
      \bigg]\,,
      \\[5mm]
    d=6\!:~~~&\displaystyle
    S_\text{ct} 
  = \frac{1}{\kappa^2} \int_{\partial M_\eps} d^d x  \sqrt{\gamma}\, \bigg[ 
      \frac{5}{L} 
      + \frac{L}{8} R 
      +\frac{L^3}{64} \bigg( R_{ij} R^{ij} - \frac{3}{10} R^2 \bigg)
      \\[5mm]
      &
     \displaystyle
     \hspace{4cm}
      - \log\rho\, \frac{L^5}{256} \bigg ( R_{ij} \square R^{ij} 
      - \frac{1}{20} R \square R 
      \\
      &
       \displaystyle
     \hspace{5cm}
      + 2 R^{ij} R^{kl} R_{ikjl} + \frac{1}{5} R R_{ij} R^{ij} - \frac{3}{100} R^3 \bigg )
            \bigg]\,,
  \end{array}
  }
\ee
where we have used $\rho = e^{-2r/L}$. The results for the six-derivative terms displayed for $d=6$ are derived in Appendix \ref{app:SixDerivatives}.

These purely gravitational counterterms reproduce results well-known in the literature, see for example \cite{deHaro:2000vlm}, but it is relevant to present them here in the context of our approach to holographic renormalization. In particular, they will appear in the following examples. 


\section{Renormalization for the ABJM model}
\label{sec:ABJMmodel}

ABJM theory \cite{Aharony:2008ug} is the $\mathcal{N} = 6$ superconformal Chern-Simons theory in $d = 3$ dimensions with gauge group $U ( N ) \times U ( N )$ and Chern-Simons levels $k$ and $-k$.
Its holographic dual is M-theory on $AdS_4 \times S^7 / \mathbb{Z}_k$.
In the limit of large t'Hooft coupling ($\lambda = N / k$), M-theory reduces to eleven dimensional supergravity on $AdS_4 \times S^7 / \mathbb{Z}_k$.
The recent paper  \cite{Freedman:2013ryh} by Freedman and Pufu explores the gauge-gravity dual description of $F$-maximization for ABJM theory on a 3-sphere using a 4-dimensional holographic dual. We will use the model of \cite{Freedman:2013ryh} as a  very simple example to illustrate our approach to holographic renormalization. 

The ABJM holographic model \cite{Freedman:2013ryh} is described by the Euclidean bulk action 
\begin{equation}
S_\textrm{bulk} = - \frac{1}{2 \kappa^2} \int_M d^3 x\, dr \sqrt{g	} \big( \mathcal{R} [ g ] - \mathcal{L}_m \big)\,,
\end{equation}
where $\kappa^2 = 8\pi G_4$ and the matter Lagrangian is
\begin{equation}
\mathcal{L}_m = 2 \sum_{a = 1}^3 \frac{\partial_\mu z^a \partial^\mu \bar{z}^a}{( 1 - z^a \bar{z}^a )^2} + V ( z, \bar{z} )\,,
~~~~~~~
V ( z, \bar{z} ) = \frac{1}{L^2} \bigg ( 6 - \sum_{a = 1}^3 \frac{4}{1 - z^a \bar{z}^a} \bigg )\,.
\end{equation}
In the Euclidean theory, the scalars $z^a$ and $\bar{z}^a$ are independent complex fields, not related by complex conjugation. However, since only products of $z^a$ and $\bar{z}^a$ appear in this Lagrangian, it is useful to define $z^a \to \frac{1}{\sqrt{2}} \left ( \chi^a + i \psi^a \right ), \bar{z}^a \to \frac{1}{\sqrt{2}} \left ( \chi^a - i \psi^a \right )$, where $\chi^a$ and $\psi^a$ are fields that can take complex values.

Under this, the matter Lagrangian becomes
\begin{equation}
\mathcal{L}_m = \sum_{a = 1}^3 \frac{\partial_\mu \chi^a \partial^\mu \chi^a + \partial_\mu \psi^a \partial^\mu \psi^a}{\left [ 1 - \frac{1}{2} ( \chi^a )^2 - \frac{1}{2} ( \psi^a )^2 \right ]^2} 
+  V\,,
~~~~~~~
V =
 \frac{1}{L^2} \bigg ( 6 - \sum_{a = 1}^3 \frac{4}{1 - \frac{1}{2} ( \chi^a )^2 - \frac{1}{2} ( \psi^a )^2} \bigg ) \,.
\end{equation}
Expanding the potential for small fields, we find
\be
  V  =  \frac{1}{L^2}  \Big(  -  6 - 2 (\chi^a\chi^a + \psi^a\psi^a ) - (\chi^a\chi^a + \psi^a\psi^a)^2 + \ldots \Big)\,,
\ee
so the six fields $\chi^a$ and $\psi^a$ all have mass $-2/L^2$. By our general discussion, this means that their asymptotic falloff is generically $e^{-r/L}$.

For simplicity, let us start out with a model with just one pair of the fields  $\chi$ and $\psi$; since the ABJM dual has the three pairs appear the same way and they do not mix, it is easy to generalize the result back to that case. Thus setting the fields with $a=2,3$ to zero, we will consider the model  described by the 
potential 
\be
 V= \frac{1}{L^2} \bigg ( - 2 - \frac{4}{1 - \frac{1}{2} \chi^2 - \frac{1}{2} \psi^2} \bigg )\,.
\ee
In the notation \reef{Sbulk}, we have scalars $\Phi^I = (\chi,\psi)$ and 
 the metric on the scalar target space is the $G_{IJ} = \left ( 1 - \frac{1}{2} \chi^2 - \frac{1}{2} \psi^2 \right )^{-2} \delta_{IJ}$ with $I,J=1,2$. 
The Hamilton-Jacobi equation \reef{eq:HJEquationFinal2} for this model is then
\begin{multline}
R + \mathcal{K} - \Big ( 1 - \frac{1}{2} \chi^2 - \frac{1}{2} \psi^2 \Big )^{-2} \gamma^{ij} ( \partial_i \chi \partial_j \chi  + \partial_i \psi \partial_j \psi ) \\
+ \left ( 1 - \frac{1}{2} \chi^2 - \frac{1}{2} \psi^2 \right )^2 \left ( p_\chi^2 + p_\psi^2 \right ) - \frac{1}{L^2} \left ( - 2 - \frac{4}{1 - \frac{1}{2} \chi^2 - \frac{1}{2} \psi^2} \right ) + 2 \frac{\partial U}{\partial r} = 0\,,
\label{eq:HJEquationABJM}
\end{multline}
where $\mathcal{K}$ is given by equation~(\ref{KKinHJ}) and the conjugate momenta $p_\chi$ and $p_\psi$ are the $\chi$ and $\psi$ derivatives of the on-shell action (\ref{pI}). We now proceed to determine the infinite counterterms for this model.

\paragraph{Step 1:}
Since we are working in $d = 3$ dimensions we need to include in our Ansatz only terms with up to two derivatives:
\begin{equation}
U = U_{(0)} + U_{(2)}\,.
\end{equation}
Terms with four or more derivatives give finite contributions to the on-shell action.

Keeping only potentially divergent contributions means that for $U_{(0)}$ we only need to consider terms up to cubic order in the scalar fields. However, we get strong constraints on the Ansatz from the symmetries of the model: it is invariant under the transformations $\chi \to - \chi$, $\psi \to - \psi$, and $\chi \leftrightarrow \psi$. With these symmetries imposed, the most general Ansatz at zero-derivative order is
\begin{equation}
U_{(0)} = - \frac{2}{L} + A ( r ) ( \chi^2 + \psi^2 )\,.
\end{equation}
The constant term is fixed from the purely gravitational calculation of Section \ref{sec:PureGravity}.
At two-derivative order, the only potentially divergent term that preserves the symmetries of the theory is purely gravitational and it was calculated in Section~\ref{sec:PureGravity}:
\begin{equation}
U_{(2)} = - \frac{L}{2} R\,.
\end{equation}
We can skip {\bf Step 2} because the model is so simple. 

\paragraph{Step 3:}
We are now able to solve Equation~(\ref{eq:HJEquationABJM}).
Keeping only zero-derivative terms and using that $\mathcal{K}_{(0)} = - \frac{3}{2} U_{(0)}^2$ from (\ref{KKinHJ}) we find that
\begin{equation}
- \frac{3}{2} U_{(0)}^2 + \left ( 1 - \frac{1}{2} \chi^2 - \frac{1}{2} \psi^2 \right )^2 ( p_{\chi (0)}^2 + p_{\psi (0)}^2 ) - V ( \chi, \psi ) + 2 \frac{\partial U_{(0)}}{\partial r} = 0\,,
\end{equation}
where,
\be
p_{\chi (0)} = \frac{\delta U_{(0)}}{\delta \chi} = 2 A \chi\,, 
~~~~~
p_{\psi (0)} = \frac{\delta U_{(0)}}{\delta \psi} = 2 A \psi\,.
\ee
Putting everything together and collecting terms that are proportional to $( \chi^2 + \psi^2 )$ gives the following differential equation for $A ( r )$:
\begin{equation}
\dot{A} + 2 A^2 + \frac{3}{L} A + \frac{1}{L^2} = 0\,.
\end{equation}
This has solution
\begin{equation}
A = - \frac{1}{2 L} + \mathcal{O} \left ( e^{- r / L} \right )\,.
\end{equation}
Since $A$ was the only unknown coefficient in the Ansatz for $U$, this concludes the calculation of the infinite contributions in the on-shell action.
Specifically, we have found that
\begin{equation}
U_{(0)} = - \frac{1}{L} \left ( 2 + \frac{1}{2} \chi^2 + \frac{1}{2} \psi^2 \right ) = - \frac{1}{L} \left ( 2 + z \bar{z} \right ) \,.
\end{equation}

\paragraph{Step 4:}
The counterterm action for the ABJM model is obtained by generalizing our result to the three flavors of $z^a$ and $\bar{z}^a$ fields:
\begin{equation}
S_{ct} = \frac{1}{\kappa^2} \int_{\partial M_\eps} d^3 x \sqrt{\gamma} \Bigg [ \frac{1}{L} \bigg ( 2 + \sum_{a = 1}^3 z^a \bar{z}^a \bigg ) + \frac{L}{2} R \Bigg ] \,.
\end{equation}
This result is in perfect agreement with the counterterm action given in equations (6.4)-(6.5) in \cite{Freedman:2013ryh}. For the applications in  \cite{Freedman:2013ryh} one further needs to use supersymmetry to determine the finite counterterms; we do not discuss this   here. 

\section{Renormalization for the FGPW model}
\label{sec:FGPW}

The FGPW model \cite{Freedman:1999gp} is the holographic dual of the single-mass limit of  $\mathcal{N}=1^*$ gauge theory in flat space. This non-conformal field theory is obtained from $\mathcal{N}=4$ SYM theory by softly breaking the supersymmetry to $\mathcal{N}=1$ as follows. In $\mathcal{N}=1$ language,  $\mathcal{N}=4$ SYM consists of a vector multiplet and three chiral multiplets. The field theory dual to FGPW is obtained by giving a mass to one of the chiral multiplets. In the UV, the conformal theory of $\mathcal{N}=4$ SYM is recovered, while in the infrared, the theory flows to a Leigh-Strassler fixed point. The holographic dual FGPW model captures the RG flow of this theory via a  flat-space sliced domain wall solution which approaches asymptotic $AdS_5$ in the UV and another  $AdS_5$ in the IR. The ratio of the AdS radii in the UV and IR  translates to the ratio of UV and IR central charges $a$ in the field theory.  More generally, the authors of \cite{Girardello:1998pd,Freedman:1999gp} derived the first version of a holographic version of the $c$-theorem.

The holographic FGPW model is described by a  $D = 4 + 1$-dimensional bulk action
\begin{equation}
S = - \frac{1}{2 \kappa^2} \int_{M} d^4 x\, dr \sqrt{g}\, \big ( \mathcal{R}[g]  - \mathcal{L}_m \big )\,,
\label{eq:FGPWaction}
\end{equation}
with matter Lagrangian given by\footnote{In the paper \cite{Freedman:1999gp}, the scalar potential $V$ is given in terms of a superpotential $W$ as
\be
   \label{VfgpwOrig}
   V_\text{FGPW} = \frac{1}{L^2} \bigg( \frac{1}{2} \bigg| \frac{\pa W}{\pa \phi_1}\bigg|^2
         + \frac{1}{2} \bigg| \frac{\pa W}{\pa \phi_3}\bigg|^2 
         - \frac{4}{3} W^2 \bigg)\,,
\ee
with 
\be
   \label{WfgpwOrig}
  W = \frac{1}{4\rho^2} 
  \Big[
     \cosh(2 \phi_1) (\rho^6 - 2) - (3 \rho^6 +2)
  \Big] 
  ~~~~\text{and}~~~~
  \rho = e^{\phi_3/\sqrt{6}}\,.
\ee
Here, we have conformed to our normalization conventions by rescaling the scalars $\phi_1 = \psi/\sqrt{2}$ and $\phi_3 = \phi/\sqrt{2}$, and taken the potential to be $V=4 V_\text{FGPW}$.
}

\begin{equation}
\mathcal{L}_m = \partial_\mu \phi \partial^\mu \phi + \partial_\mu \psi \partial^\mu \psi + V ( \phi, \psi ) = \dot{\phi}^2 + \dot{\psi}^2 + \gamma^{ij} \partial_i \phi \partial_j \phi + \gamma^{ij} \partial_i \psi \partial_j \psi + V ( \phi, \psi ) \,.
\end{equation}
The scalars $\psi$ and $\phi$ are dimension $\Delta_\psi = 3$ and  $\Delta_\phi = 2$ fields dual to the fermion and scalar mass deformations of $\mathcal{N}=4$ SYM. They approach zero near the UV boundary as 
\be
  \label{FGPWfalloff}
 \psi \sim \psi_0\, e^{-r/L} 
 ~~~~\text{and}~~~~ 
 \phi \sim  (\phi_0 r + \tilde{\phi}_0 )\, e^{-2r/L}\,,
\ee
 as $r\to \infty$. For the purpose of holographic renormalization, we only need to keep the  terms in the potential that can give divergent terms in this limit, so we expand the potential in small fields to find 
\begin{equation}
\label{Vfgpw}
V ( \phi, \psi ) = \frac{1}{L^2} \left ( - 12 - 4 \phi^2 - 3 \psi^2 + c \psi^4 
+ \ldots \right )\,.
\end{equation}
The masses of the scalars, $m_\psi^2 = -3/L^2$ and $m_\phi^2 = -4/L^2$, are directly related to the scaling dimensions $\Delta_\psi = 3$ and  $\Delta_\phi = 2$ via $m_I^2 L^2 = \Delta_I ( \Delta_I - 4 )$.

The actual FGPW model has $c = 1$ 
 in \reef{Vfgpw}, but here we keep the coefficients general. This will serve to illustrate how the counterterms carry information that is specifically dependent on coefficients in the scalar potential; i.e.~one should in general expect model-dependent terms in the counterterm action.  

The HJ equation \reef{eq:HJEquationFinal2} for the FGPW model takes the form
\begin{equation}
~R[\gamma] +  \mathcal{K}  + p_\phi^2 +  p_\psi^2 
- \gamma^{ij} \partial_i \phi \partial_j \phi 
- \gamma^{ij} \partial_i \psi \partial_j \psi  - V ( \phi, \psi ) + 2 \frac{\partial U}{\partial r} = 0\,.~
\label{eq:HJEquationFinalFGPW}
\end{equation}
with $\mathcal{K}$ defined in \reef{KKinHJ} and momenta
\begin{equation}
p_\phi = \frac{\delta U}{\delta \phi} \qquad p_\psi = \frac{\delta U}{\delta \psi}\,.
\label{eq:JacobiMomentaPhiFGPW}
\end{equation}

Since we are working in $d = 4$ dimensions we need to keep terms with up to four derivatives, so we write
\begin{equation}
U = U_{(0)} + U_{(2)} + U_{(4)}\,.
\end{equation}
We now proceed with solving for the divergent terms of the on-shell action following the algorithmic procedure described in Section \ref{s:algo}:
\paragraph{Step 1:} We begin by writing the most general Ansatz for each $U_{(i)}$. 
We only keep terms that can give divergent contributions. With the scalar falloffs \reef{FGPWfalloff} and each inverse metric giving $e^{-2r}$, the most general Ansatz at 0th order is
\begin{equation}
U_{(0)} =  - \frac{3}{L} + A_1 \psi + A_2 \phi + A_3 \psi^2 + A_4 \phi \psi + A_5 \psi^3 + A_6 \phi^2 + A_7 \phi \psi^2 + A_8 \psi^4\,,
\label{eq:FGPWU0}
\end{equation}
where the constant term is fixed by the purely gravitational analysis in Section \ref{sec:PureGravity}. Each of the coefficients $A_i$  is considered a function of $r$.

At order 2 we use the Ansatz
\begin{equation}
\label{eq:FGPWU2}
U_{(2)} = - \frac{L}{4} R + B_1 R \psi + B_2 R \phi + B_3 R \psi^2 + B_4 \psi \square \psi\,.
\end{equation}
We did not include $( \partial \psi )^2$, since it is equivalent to $\psi \square \psi$  after partial integration.

At order 4, the only option are the purely gravitational terms we have already solved, so we have
\begin{equation}
U_{(4)} = - \frac{L^2 r}{8} \left ( R_{ij} R^{ij} - \frac{1}{3} R^2 \right ) \,.
\end{equation}
Since the full FGPW model \reef{VfgpwOrig}-\reef{WfgpwOrig} is symmetric under $\psi \to -\psi$, we can immediately set the following coefficients in the Ansatz to zero:
\be
  A_1 =A_4 =A_5=B_1=0\,.
\ee

\paragraph{Step 2:} At the leading order, the conjugate momenta obtained from (\ref{eq:HamiltonMomenta}) must agree with those in \reef{eq:JacobiMomentaPhiFGPW}.
From (\ref{eq:HamiltonMomenta}), we have
\begin{equation}
p_\phi = \dot{\phi} \qquad p_\psi = \dot{\psi}\,,
\end{equation}
and via \reef{FGPWfalloff} this gives
\begin{equation}
  p_\phi = - \frac{2}{L} \Big ( 1 - \frac{L}{2 r} \Big ) \phi + \mathcal{O}  \big( e^{- 2 r / L} / r \big ) \,,
  \qquad 
  p_\psi = - \frac{1}{L} \psi + \mathcal{O}\big ( e^{-3 r / L} \big )\,.
\label{eq:FGPWMomentumAsymptotics}
\end{equation}
On the other hand (\ref{eq:JacobiMomentaPhiFGPW}) gives 
\begin{equation}
\begin{aligned}
\label{fgpwp0}
p_{\phi (0)}  = \frac{\delta U_{(0)}}{\delta \phi} = A_2  
+ 2 A_6 \phi + A_7 \psi^2\,, 
~~~~~~
p_{\psi (0)}  = \frac{\delta U_{(0)}}{\delta \psi} = 
 2 A_3 \psi 
+ 2 A_7 \phi \psi + 4 A_8 \psi^3\,.
\end{aligned}
\end{equation}
Comparing  \reef{eq:FGPWMomentumAsymptotics} to terms in \reef{fgpwp0} at similar orders, we can directly infer that some of the  coefficients $A_i$ must vanish:
\be
 A_2 = A_7 = 0\,.
\ee
Furthermore, we learn that $A_3 = - \frac{1}{2L}$ and $A_6 = - \frac{1}{L} \big ( 1 - \frac{L}{2 r} \big )$. However, let us leave $A_3$ and $A_6$ unfixed for now  for the purpose of illustrating how they are fixed using the HJ equation. 

\paragraph{Step 3:} We proceed to solve the HJ equation \reef{eq:HJEquationFinalFGPW}. We start from the terms at 0th order. Keeping only  terms without spatial derivatives and using $\mathcal{K}_{(0)} = - \frac{4}{3} U_{(0)}^2$ from \reef{K0} we find that
\begin{equation}
- \frac{4}{3} U_{(0)}^2 + p_{\phi (0)}^2 + p_{\psi (0)}^2 - V ( \phi , \psi ) + 2 \frac{\partial U_{(0)}}{\partial r} = 0\,.
\end{equation}
To solve this, we set the coefficient of each combination of fields to zero. For example, collecting the terms proportional to $\psi^2$ gives 
\be
  \dot{A}_3 + \frac{4}{L} A_3 + 2 A_3^2 + \frac{3}{2 L^2} = 0
  ~~~~\implies~~~~
  A_3 = - \frac{1}{2 L} + \mathcal{O} \big( e^{-2 r / L} \big)\,.
\ee
This is the solution for $A_3$ we anticipated from comparing \reef{eq:FGPWMomentumAsymptotics} and \reef{fgpwp0}.

Similarly, one finds
\be
\begin{array}{lrlll}
\textrm{$\phi^2$-terms:} ~
& \displaystyle \dot{A}_6 + \frac{4}{L} A_6 + 2 A_6^2 + \frac{2}{L^2} = 0 
& \implies 
& \displaystyle A_6 = - \frac{1}{L} + \frac{1}{2 r} + \mathcal{O} \Big ( \frac{L^2}{r^2} \Big ) \,,
\\[3mm]
\textrm{$\psi^4$-terms:} 
&\displaystyle \dot{A}_8 - \frac{1}{6 L^2} \left ( 1 + 3 c \right) = 0 
& \implies 
& \displaystyle A_8 = \frac{1}{6 L^2} \left ( 1 + 3 c \right) r + \mathcal{O} \big ( 1 \big )\,.
\end{array}
\ee
Terms proportional to $\phi \psi^2$  vanish directly; had we had a term $b \phi \psi^2$ in the expansion of the scalar potential, the HJ equation would have shown that $b \ne 0$ is not consistent with the EOM. 

Having calculated all the unknown coefficients in the $U_{(0)}$ Ansatz, let us write down the final result (with $r = - \frac{L}{2} \log \rho$):
\begin{equation}
\label{U0fgpw}
U_{(0)} = - \frac{1}{L} \left [ 3 + \bigg ( 1 + \frac{1}{\log \rho} \bigg ) \phi^2 + \frac{1}{2} \psi^2 + \frac{1}{12} \big ( 1 + 3 c \big ) \psi^4 \log \rho \right ]\,.
\end{equation}
We can identify each of the contributions. 
The first one is related to the cosmological constant and it is fixed for all models in $D = 4 + 1$ dimensions, as we saw in the pure gravity case in Section \ref{sec:PureGravity}.
The terms that are quadratic in the fields are uniquely fixed by the mass terms in the scalar potential and are as such universal for all models. Finally, the $\psi^4$-terms are clearly model-dependent, as can be seen from the explicit dependence on $c$.

With the 0th order result in hand, we are now able to continue solving HJ equation for the two-derivative terms.
Keeping only such terms from equation~(\ref{eq:HJEquationFinalFGPW}) gives
\begin{equation}
\label{HJ2FGPW}
R - \frac{8}{3} U_{(0)} \left ( U_{(2)} - \frac{1}{2} Y_{(2)} \right ) + 2 p_{\phi (0)} p_{\phi (2)} + 2 p_{\psi (0)} p_{\psi (2)} - \gamma^{ij} \partial_i \phi \partial_j \phi - \gamma^{ij} \partial_i \psi \partial_j \psi + 2 \frac{\partial U_{(2)}}{\partial r} = 0\,,
\end{equation}
where we used $\mathcal{K}_{(2)}$ from \reef{K2K4}.  $U_{(0)}$, $p_{\phi (0)}$ and $p_{\psi (0)}$ are known from \reef{fgpwp0} and \reef{U0fgpw}, while we calculate $p_{\phi (2)}$ and $p_{\psi (2)}$, and $Y_{(2)}$ from the Ansatz \reef{eq:FGPWU2} for $U_{(2)}$:
\begin{equation}
\begin{aligned}
p_{\phi (2)} & = \frac{\delta U_{(2)}}{\delta \phi} = B_2 R\,, \\
p_{\psi (2)} & = \frac{\delta U_{(2)}}{\delta \psi} =  
2 B_3 R \psi + 2 B_4 \square \psi\,, \\
Y_{(2) ij} & = \frac{\delta U_{(2)}}{\delta \gamma^{ij}} = 
- \frac{L}{4} R_{ij} 
+ B_2  R_{ij} \phi + B_3 R_{ij} \psi^2 + B_4 \psi \nabla_i \nabla_j \psi \,,
\end{aligned}
\end{equation}
where we are dropping total derivatives. The result for $Y_{(2) ij}$ implies $Y_{(2)} = U_{(2)}$. 
In the HJ equation \reef{HJ2FGPW}, we organize the terms according to the field monomials and set the coefficients of divergent terms to zero. The terms simply proportional to  $R$ directly vanish because we have already solved the purely gravitational part of the problem. The remaining terms allow us to solve for the coefficients $B_{2,3,4}$ :
\be
\begin{array}{lrlll}
\textrm{$R \phi$-terms:} 
& \displaystyle  \dot{B}_2 + \frac{1}{r} B_2 = 0 
& \implies 
& \displaystyle B_2 = \mathcal{O}  \Big ( \frac{1}{r} \Big ) \,,
\\[3mm]
\textrm{$R \psi^2$-terms:} 
& \displaystyle \dot{B}_3 - \frac{1}{12} = 0 
& \implies 
& \displaystyle B_3 = \frac{1}{12} r + \mathcal{O} \big ( 1 \big ) \,,
\\[3mm]
\textrm{$\psi \square \psi$-terms:} 
& \displaystyle \dot{B}_4 + \frac{1}{2} = 0 
& \implies 
& \displaystyle B_4= - \frac{1}{2} r + \mathcal{O} \big ( 1 \big )\,.
\end{array}
\ee
As in the zero weight case the subleading terms related to integration constants are not important because they lead to finite contributions to the action.
The final expression for $U_{(2)}$ is then 
\begin{equation}
U_{(2)} = - L \left [ \frac{1}{4} R - \frac{1}{4} \psi \left ( \square - \frac{1}{6} R \right ) \psi \log \rho \right ]\,.
\end{equation}
The first term is purely gravitational. The second term is independent of details of the  higher order terms in the potential and thus fixed for all models that contain a scalar with $m^2 L^2 = - 3$.
Finally, notice that the combination of the Laplace operator $\square$ and the Ricci scalar $R$ that appears in the last term is proportional, up to an overall constant to the conformal Laplacian.

\paragraph{Step 4:}
We have now fully determined the counterterm action necessary to cancel the divergences of the on-shell action.
In particular we will have $S_\text{ct} = - \frac{1}{\kappa^2} \int d^4 x \sqrt{\gamma} \,U$ and therefore,
\begin{multline}
S_\text{ct} = \frac{1}{\kappa^2} \int_{\partial M_\eps} d^4 x \sqrt{\gamma} \Bigg \{ \frac{1}{L} \left [ 3 + \left ( 1 + \frac{1}{\log \rho} \right ) \phi^2 + \frac{1}{2} \psi^2 + \frac{1}{12} \left ( 1 + 3 c \right ) \psi^4 \log \rho \right ] \\
+ L \left [ \frac{1}{4} R - \frac{1}{4} \psi \left ( \square - \frac{1}{6} R \right ) \psi \log \rho \right ] - \frac{1}{16} L^3 \left ( R_{ij} R^{ij} - \frac{1}{3} R^2 \right ) \log \rho \Bigg \}\,.
\label{eq:FGPWct}
\end{multline}
This is our final result for the FGPW model.

As a test, we have calculated the one-point functions of the QFT operators that are dual to the fields of the FGPW model.
The one-point function of the operator dual to field $\phi^I$ will be given by\footnote{
In the special case where $\Delta_I = d / 2$ the one-point function has an extra factor of $\log \rho$.}
\begin{equation}
\langle O_{\phi^I} \rangle = - \lim_{\rho \to 0} \frac{\rho^{- \Delta_I / 2}}{\sqrt{\gamma}} \frac{\delta S_\textrm{ren}}{\delta \phi^I}\,,
\end{equation}
where the regularized action (ignoring possible finite counterterms) is
\begin{equation}
S_\textrm{reg} =  
S_\textrm{bulk} + S_\textrm{GH} + S_\textrm{ct} 
 \, .
\end{equation}
In order to check that the expressions obtained are indeed finite, one must impose the equations of motion on the coefficients in the Fefferman-Graham expansion of the fields. We find that with our infinite counterterms, all three  one-point functions in FGPW are indeed finite.
Details are presented in Appendix~\ref{app:1ptFunctions}.


\section{Renormalization of a dilaton-axion model}
\label{sec:dilaton}

In this section we present the procedure of renormalization of a dilaton-axion model.
The purpose of this example is to illustrate how the procedure for holographic renormalization applies to theories that include marginal scalars.
Specifically, we examine the renormalization of the dilaton-axion model previously studied in \cite{Papadimitriou:2011qb}: the 5d bulk action is
\begin{equation}
S_\textrm{bulk} = - \frac{1}{2 \kappa^2} \int_M d^4 x\, dr \sqrt{g} \big( \mathcal{R} [ g ] - \mathcal{L}_m \big)\,,
\label{eq:DilatonAction}
\end{equation}
with
\begin{equation}
\mathcal{L}_m =  \partial_\mu \varphi \partial^\mu \varphi + Z ( \varphi ) \partial_\mu \chi \partial^\mu \chi - \frac{12}{L^2}\,.
\end{equation}

The fields $\varphi$ and $\chi$ are massless and therefore correspond to marginal QFT operators with scaling dimension $\Delta = 4$. $Z$ denotes an arbitrary function of the dilaton field $\varphi$. 
Near the asymptotic boundary, these scalars generically do not vanish but instead approach a finite value.
In particular, their asymptotic behavior is given by
\be
\varphi ( x, r )  =  \varphi_{(0)} ( x ) + \mathcal{O} \Big ( e^{-2 r / L} \Big )\,,
~~~~~~
\chi ( x, r )  =  \chi_{(0)} ( x ) + \mathcal{O} \Big ( e^{-2 r / L} \Big )\,.
\label{eq:DilatonAxionAsymptotics}
\ee
As a consequence, we cannot regard the effective action as a power-expansion in these fields, as higher powers are not suppressed. Instead, we will take the Ansatz to involve general functions of $\varphi$ and $\chi$.

By defining the field $\Phi$ to be $\Phi = ( \varphi, \chi )$ and the K\"{a}hler metric to be $G = \left ( \begin{matrix} 1 & 0 \\ 0 & Z ( \varphi ) \end{matrix} \right )$, we conclude that the HJ Equation~(\ref{eq:HJEquationFinal2}) now becomes
\begin{equation}
R[\gamma] + \mathcal{K} + p_\varphi^2 + \frac{1}{Z ( \varphi )} p_\chi^2 - \gamma^{ij} \partial_i \varphi \partial_j \varphi - Z ( \varphi ) \gamma^{ij} \partial_i \chi \partial_j \chi + \frac{12}{L^2} + 2 \frac{\partial U}{\partial r} = 0\,.
\label{eq:HJEquationDilatonAxion}
\end{equation}
The momenta are defined, in the usual way \reef{pI}, as derivatives of $U$.

Let us now examine step-by-step the procedure introduced in the previous sections and spot any important differences.

\paragraph{Step 1:}
With $d = 4$, we need to keep terms with up to four derivatives:
\begin{equation}
U = U_{(0)} + U_{(2)} + U_{(4)}\,.
\end{equation}
Taking into account that any possible function of the fields could give divergent contributions in the on-shell action we write the following Ansatz for the zero, two and four derivative parts of $U$ respectively:
\bea
U_{(0)}&\!=\!& A ( \varphi, \chi, r )\,, \\[2mm]
U_{(2)} &\!=\!& B_0 R + B_1 ( \nabla \varphi ) \cdot ( \nabla \chi ) + B_2 ( \nabla \varphi )^2 + B_3 (\nabla \chi )^2\,, \\[2mm]
\nonumber
U_{(4)} &\!=\!& C_1 R^2 + C_2 R_{ij} R^{ij} + C_3 R \square \varphi + C_4 R \square \chi + C_5 R ( \nabla \varphi )^2 + C_6 R ( \nabla \chi )^2 + C_7 R ( \nabla \varphi ) \cdot ( \nabla \chi ) 
\\
\nonumber
&&+ C_8 R^{ij} \nabla_i \varphi \nabla_j \varphi + C_9 R^{ij} \nabla_i \chi \nabla_j \chi + C_{10} R^{ij} \nabla_i \varphi \nabla_j \chi + C_{11} ( \square \varphi )^2 + C_{12} ( \square \chi )^2 
\\
&&+ C_{13} \square \varphi \square \chi + C_{14} \nabla_i \nabla_j \varphi \nabla^i \nabla^j \varphi + + C_{15} \nabla_i \nabla_j \chi \nabla^i \nabla^j \chi + + C_{16} \nabla_i \nabla_j \varphi \nabla^i \nabla^j \chi 
\\
\nonumber
&&+ C_{17} \square \varphi ( \nabla \varphi )^2 + C_{18} \square \chi ( \nabla \chi )^2 + C_{19} \square \varphi ( \nabla \chi )^2 + C_{20} \square \varphi ( \nabla \varphi ) \cdot ( \nabla \chi ) 
\\
\nonumber
&&+ C_{21} \square \chi ( \nabla \varphi )^2 + C_{22} \square \chi ( \nabla \varphi ) \cdot ( \nabla \chi ) + C_{23} \left ( ( \nabla \varphi )^2 \right )^2 + C_{24} \left ( ( \nabla \chi )^2 \right )^2 
\\
\nonumber
&&+ C_{25} ( \nabla \varphi )^2 ( \nabla \chi )^2 + C_{26} ( ( \nabla \varphi ) \cdot ( \nabla \chi ) )^2 + C_{27} ( \nabla \varphi )^2 ( \nabla \varphi ) \cdot ( \nabla \chi ) + C_{28} ( \nabla \chi )^2 ( \nabla \varphi ) \cdot ( \nabla \chi )\,.
\eea
The coefficients $A$, $B_i$ and $C_i$ are all considered functions of the radial coordinate $r$ as well as the fields $\varphi$ and $\chi$.
We have omitted terms that up to total derivatives can be decomposed to the ones already included.
For example, since $B \square \varphi = \nabla_i ( B \nabla^i \varphi ) - \partial_\varphi B ( \nabla \varphi )^2 - \partial_\chi B (\nabla \varphi ) \cdot ( \nabla \chi )$, such a term can be absorbed in $B_1$ and $B_2$, so it is redundant to include it in the Ansatz.

\paragraph{Step 2:}
We use equation~(\ref{eq:HamiltonMomenta}) and the asymptotic behavior of the fields~(\ref{eq:DilatonAxionAsymptotics}) to determine the leading behavior of $p_{\varphi}$ and $p_{\chi}$ to be
\begin{equation}
p_\varphi = \dot{\varphi} = \mathcal{O} \Big ( e^{- 2 r / L} \Big )\,,
\qquad
p_\chi = Z ( \varphi ) \dot{\chi} = \mathcal{O} \Big ( e^{- 2 r / L} \Big )\,.
\end{equation}
On the other hand, our Ansatz for $U_{(0)}$ gives
\begin{equation}
p_{\varphi (0)}  = \frac{\delta U_{(0)}}{\delta \varphi} = \partial_\varphi A \,,
~~~~~~
p_{\chi (0)} = \frac{\delta U_{(0)}}{\delta \chi} = \partial_\chi A\,.
\end{equation}
By comparing the two sets of expressions for the momenta, we understand that the coefficient $A$ can neither depend on $\varphi$ nor $\chi$, and thus $p_{\varphi (0)}$ and $p_{\chi (0)}$ vanish.
This leaves $U_{(0)}$ to be purely gravitational and thus we can use directly our result from Section~\ref{sec:PureGravity}:
\begin{equation}
U_{(0)} = - \frac{3}{L}\,.
\end{equation}

\paragraph{Step 3:}
We now proceed to solve HJ equation and determine the unknown coefficients of our Ansatz. 
Since the zero-derivatives contribution has already been fixed, we start our analysis with the two-derivative terms.
At this order, the HJ equation simplifies to
\begin{equation}
R - \frac{8}{3} U_{(0)} \left ( U_{(2)} - \frac{1}{2} Y_{(2)} \right ) - ( \nabla \varphi )^2 - Z ( \varphi ) ( \nabla \chi )^2 + 2 \frac{\partial U_{(2)}}{\partial r} = 0\,
\end{equation}
 using $p_{\varphi (0)}=p_{\chi (0)}=0$. Here, $Y_{(2)} = \gamma^{ij} Y_{(2) ij}$ is the trace of the tensor
\be
\begin{split}
Y_{(2) ij} = \frac{\delta U_{(2)}}{\delta \gamma^{ij}} 
=&\, B_0 R_{ij} - \nabla_i \nabla_j B_0 + \square B_0 \gamma_{ij} + \frac{1}{2} B_1 \nabla_i \varphi \nabla_j \chi \\
&+ \frac{1}{2} B_1 \nabla_i \chi \nabla_j \varphi + B_2 \nabla_i \varphi \nabla_j \varphi + B_3 \nabla_i \chi \nabla_j \chi\,.
\end{split}
\ee
After plugging everything into the HJ equation, one uses partial integration to eliminate terms that were not in our original Ansatz and therefore were not independent. 
Demanding that the coefficient of each independent term in the resulting HJ equation is zero, one finds that the two-derivative contribution to the on-shell action is
\begin{equation}
U_{(2)} = - \frac{L}{4} \big [ R - ( \nabla \varphi )^2 - Z ( \varphi ) ( \nabla \chi )^2 \big ]\,.
\end{equation}

For terms with four spatial derivatives equation~(\ref{eq:HJEquationDilatonAxion}) simplifies to
\begin{multline}
- \frac{8}{3} U_{(0)} \left ( U_{(4)} - \frac{1}{2} Y_{(4)} \right ) + 4 Y_{(2) ij} Y_{(2)}^{ij} - \frac{4}{3} \left ( U_{(2)} - \frac{1}{2} Y_{(2)} \right )^2 - Y_{(2)}^2 \\
+ p_{\varphi (2)}^2 + \frac{1}{Z ( \varphi )} p_{\chi (2)}^2 + 2 \frac{\partial U_{(2)}}{\partial r} = 0\,.
\end{multline}
The canonical momenta that appear in this equation are
\begin{equation}
\begin{aligned}
p_{\varphi (2)} & = \frac{\delta U_{(2)}}{\delta \varphi} = - \frac{L}{2} \square \varphi + \frac{L}{4} Z' ( \varphi ) ( \nabla \chi )^2 \\
p_{\chi (2)} & = \frac{\delta U_{(2)}}{\delta \chi} = - \frac{L}{2} \square \chi - \frac{L}{2} Z' ( \varphi ) ( \nabla \varphi ) \cdot ( \nabla \chi )\,.
\end{aligned}
\end{equation}
It is useful to notice that
\begin{equation}
Y_{(4)} = \gamma^{ij} \frac{\delta U_{(4)}}{\delta \gamma^{ij}} = 2 U_{(4)} + \textrm{total derivatives}\,,
\end{equation}
and the complicated tensor $Y_{(4) ij}$ is not needed for the calculation.
The total derivatives of $Y_{(4)}$ will not contribute to HJ equation since they are multiplied by $U_{(0)}$, which is a constant, and total derivatives can be dropped by the equation.

Demanding that the different kinds of terms that appear in the four-derivative equation vanish independently yields the following solution for $U_{(4)}$:
\begin{multline}
U_{(4)} = \frac{L^3}{16} \left [ R_{ij} R^{ij} - \frac{1}{3} R^2 - 2 \left ( R^{ij} - \frac{1}{3} R \gamma^{ij} \right ) ( \nabla_i \varphi \nabla_j \varphi + Z ( \varphi ) \nabla_i \chi \nabla_j \chi )  \right . \\
+ \left ( \square \varphi - \frac{1}{2} Z' ( \varphi ) ( \nabla \chi )^2 \right )^2 + Z ( \varphi ) \left ( \square \chi + \frac{Z' ( \varphi )}{Z ( \varphi )} ( \nabla \varphi ) \cdot ( \nabla \chi ) \right )^2 \\
\left . + \frac{2}{3} \left ( ( \nabla \varphi )^2 + Z ( \varphi ) ( \nabla \chi )^2 \right )^2 + 2 Z ( \varphi ) \left ( \left ( ( \nabla \varphi ) \cdot ( \nabla \chi ) \right )^2 - ( \nabla \varphi )^2 ( \nabla \chi )^2 \right ) \right ] \log \rho \,.
\end{multline}

\paragraph{Step 4:}
This concludes the calculation of the counterterms that cancel the infinities of the on-shell action for the dilaton-axion model.
For completeness, let us write down the general result.
\begin{multline}
S_{ct} = \frac{1}{\kappa^2} \int_{\partial M_\eps} d^4 x \sqrt{\gamma} \left \{ \frac{3}{L} + \frac{L}{4} \left [ R - ( \nabla \varphi )^2 - Z ( \varphi ) ( \nabla \chi )^2 \right ] \right .\\
- \frac{L^3}{16} \left [ R_{ij} R^{ij} - \frac{1}{3} R^2 - 2 \left ( R^{ij} - \frac{1}{3} R \gamma^{ij} \right ) ( \nabla_i \varphi \nabla_j \varphi + Z ( \varphi ) \nabla_i \chi \nabla_j \chi )  \right . \\
+ \left ( \square \varphi - \frac{1}{2} Z' ( \varphi ) ( \nabla \chi )^2 \right )^2 + Z ( \varphi ) \left ( \square \chi + \frac{Z' ( \varphi )}{Z ( \varphi )} ( \nabla \varphi ) \cdot ( \nabla \chi ) \right )^2 \\
\left . \left . + \frac{2}{3} \left ( ( \nabla \varphi )^2 + Z ( \varphi ) ( \nabla \chi )^2 \right )^2 + 2 Z ( \varphi ) \left ( \left ( ( \nabla \varphi ) \cdot ( \nabla \chi ) \right )^2 - ( \nabla \varphi )^2 ( \nabla \chi )^2 \right ) \right ] \log \rho \right \}\,.
\label{eq:DilatonCT}
\end{multline}
This result for the counterterms agrees with the one found by a more complicated route in \cite{Papadimitriou:2011qb}.

\section{Discussion}
We have presented a simple implementation of the Hamiltonian approach to holographic renormalization. The idea of using the Hamilton-Jacobi equation is not new, but we hope that our presentation and algorithm makes the method more accessible and useful for others to use. For our own purposes, it has shown great value in the application to the holographic renormalization of a 10 scalar model dual to $\mathcal{N}=1^*$ gauge theory on $S^4$, an analysis that will be presented elsewhere \cite{Bobev:2016nua}.

Determining the infinite counterterms is typically only one part of holographic renormalization. One often needs the finite counterterms too, but just as in standard quantum field theory, this typically amounts to being a scheme-dependent question. However, in the presence of supersymmetry, one can fix the finite counterterms to be compatible with the supersymmetries in the problem. In the case of flat-sliced domain walls, this can be done using the Bogomolnyi-trick of writing the bulk action in terms of sums of squares that each vanish on the BPS equations. This rewriting requires a partial integration that leaves a boundary term that exactly becomes the counterterm action and encodes both infinite and finite counterterms. In the case of non-flat slicing, one can then argue that the universality of the counterterms allows one to pick the finite counterterms of the flat-space Bogomolnyi boundary term and use them in conjunction with the more general infinite counterterms discussed in this paper. This has worked successfully in several cases, for example  \cite{Freedman:2013ryh} and \cite{Bobev:2013cja}. The prescriptions does, however, have a bit of an ad hoc feel to it and it would be interesting to understand better the relationship between the BPS equations for curved domain walls and how/if they can be used to determine directly the infinite and finite counterterms.

\section*{Acknowledgements}
We are grateful to Nikolay Bobev, Dan Freedman, Finn Larsen, Ioannis Papadimitriou, and Silviu Pufu for useful discussions. Some calculations were done using Mathematica and the package {\tt xAct}. HE and MH are supported in part by NSF CAREER Grant PHY-0953232. HE is a Cottrell Scholar of the Research Corporation for Science Advancement. MH also acknowledges a Fulbright Fellowship by the Institute of International Education.


\appendix

\section{Some useful formulas}
\label{app:formulas}
We present here a list of formulas that are useful to computing the metric variations of various contractions of curvature tensors:
\bea
 \label{VarRule1}
 \int d^d x \sqrt{\gamma}\, X \frac{\d R}{\d \gamma^{ij}(y)} 
 \!\!&=&\!\! \sqrt{\gamma}\, \Big (  R_{ij} X + (\Box X) \gamma_{ij} - \nabla_i \nabla_j X \Big)\,,
\\[3mm] 
\label{VarRule2}
 \int d^d x \sqrt{\gamma}\, X \frac{\d (R_{kl} R^{kl})}{\d \gamma^{ij}(y)} 
 \!\!&=&\!\!  \sqrt{\gamma}\, \Big (  2 R_{ik}R^{k}{}_j X 
 +  \nabla_k \nabla_l (X R^{kl}) \gamma_{ij} 
 +  \Box (X R_{ij}) 
 -2   \nabla^k \nabla_i (X R_{kj}) \Big)\,,
\\[3mm] 
  \label{VarRule3}
  \int d^d x \sqrt{\gamma}\, X \frac{\d \tensor{R}{^k_{mln}}}{\d \gamma^{ij}(y)} 
   \!\!&=&\!\! \sqrt{\gamma} \bigg ( - \frac{1}{2} \nabla_m \nabla_l X \gamma_{in} \delta_j^k - \frac{1}{2} \nabla_n \nabla_l X \gamma_{jm} \delta_i^k + \frac{1}{2} \nabla^k \nabla_l X \gamma_{im} \gamma_{jn} \bigg )\,,
\\[3mm] 
  \nonumber
  \int d^d x \sqrt{\gamma}\, X \frac{\d \square Y}{\d \gamma^{ij}(y)} 
   \!\!&=&\!\! \sqrt{\gamma} \bigg ( X \nabla_i \nabla_j Y + \nabla_i ( X \nabla_j Y) - \frac{1}{2} \nabla_k ( X \nabla^k Y ) \gamma_{ij} \bigg ) + \int d^d x \sqrt{\gamma}\, \square X \frac{\d Y}{\d \gamma^{ij}(y)}\,.\\
  \label{VarRule4}
\eea
The fields on the RHS of these equations depend on $y$.

\section{Six derivative counterterms for pure gravity}
\label{app:SixDerivatives}

In $d = 6$ dimensions one needs to consider counterterms with up to six derivatives.
For the pure gravity case, the six-derivative Ansatz is given by equation~(\ref{eq:SixDerivativeAnsatz}).
In this Ansatz, it is possible to include terms with contractions of two or three Riemann tensors, but it is easy to show that the coefficients of such terms will be zero.

The HJ equation at six-derivative order becomes
\begin{equation}
\mathcal{K}_{(6)} + 2 \frac{\partial U_{(6)}}{\partial r} = 0\,.
\end{equation}
The total derivatives of $Y_{(4) ij}$ that appear in  $\mathcal{K}_{(6)}$ are now important because they are multiplied by the non-constant $Y_{(2) ij} = B R_{ij}$.
In particular, we have that
\begin{equation}
Y_{(4)ij} = C_1 \bigg ( 2 R^{kl} R_{ikjl} + \frac{1}{2} \square R \gamma_{ij} + \square R_{ij} - \nabla_i \nabla_j R \bigg ) + C_2 \bigg ( 2 R R_{ij} + 2 \square R \gamma_{ij} - 2 \nabla_i \nabla_j R \bigg )\,.
\end{equation}
The coefficients $B$ and $C_{1,2}$ are those calculated in Section~\ref{sec:PureGravity}.
Additionally, in the product $Y_{(2) ij} \tensor{Y}{_{(4)}^{ij}}$, terms proportional to $R^{ij} \nabla_i \nabla_j R$ can be changed to $R \nabla_i \nabla_j R^{ij} = \frac{1}{2} R \square R$ by adding appropriate total derivatives and using the Bianchi identity.
Finally, by using the variation rules of Appendix~\ref{app:formulas}, one realizes that $Y_{(6)} = 3 U_{(6)}$ up to total derivative terms that can be ignored because $Y_{(6)}$ is only multiplied by the constant $U_{(0)}$.
Putting everything together and demanding that the coefficient of each of the independent terms is zero gives differential equations for the coefficients $D_{1,2,3,4,5,6}$:
\be
\begin{array}{lr}
\textrm{$R^3$-terms:} 
& \displaystyle \dot{D}_1 + \frac{d-6}{L} D_1 - \frac{d L^4}{16 ( d - 1 )^2 (d - 2)^3} = 0 \,,
\\[3mm]
\textrm{$R R_{ij} R^{ij}$-terms:} 
& \displaystyle  \dot{D}_2 + \frac{d - 6}{L} D_2 + \frac{L^4}{4 ( d - 1 ) ( d - 2 )^2 ( d - 4 )} = 0\,, 
\\[3mm]
\textrm{$\tensor{R}{_i^j} \tensor{R}{_j^k} \tensor{R}{_k^i}$-terms:} 
& \displaystyle \dot{D}_3 + \frac{d - 6}{L} D_3 = 0 \,,
\\[3mm]
\textrm{$R^{ij} R^{kl} R_{ikjl}$-terms:} 
& \displaystyle \dot{D}_4 + \frac{d - 6}{L} D_4 + \frac{2 L^4}{( d - 2)^3 ( d - 4 )} = 0\,,
\\[3mm]
\textrm{$R \square R$-terms:} 
& \displaystyle \dot{D}_5 + \frac{d - 6}{L} D_5 - \frac{L^4}{4 ( d - 1 )( d - 2)^3 ( d - 4 )} = 0\,,
\\[3mm]
\textrm{$R_{ij} \square R^{ij}$-terms:} 
& \displaystyle \dot{D}_6 + \frac{d - 6}{L} D_6 + \frac{L^4}{(d - 2)^3 ( d - 4 )} = 0 \,.
\end{array}
\ee
Keeping only divergent contributions from the solutions of these equations, we obtain the result (for $d = 6$)
\begin{equation}
U_{(6)} = - \frac{L^4 r}{128} \bigg ( R_{ij} \square R^{ij} - \frac{1}{20} R \square R + 2 R^{ij} R^{kl} R_{ikjl} + \frac{1}{5} R R_{ij} R^{ij} - \frac{3}{100} R^3 \bigg )\,.
\end{equation}

\section{One-point functions}
\label{app:1ptFunctions}

In this appendix we calculate the one-point functions for the quantum field theory operators dual to the fields of the FGPW model and explicitly check that the counterterm contributions cancel the divergences that come from the bulk action.
One may consider three different one-point functions, $\langle O_\phi \rangle$, $\langle O_\psi \rangle$ and $\langle T_{ij} \rangle$,  where the QFT operators $O_{\phi / \psi}$ are dual to the bulk fields $\phi / \psi$ respectively and the QFT energy-momentum tensor $T_{ij}$ is dual to the metric $\gamma_{ij}$.

These one point functions can be calculated by variations of the renormalized action
\begin{equation}
S_\textrm{ren} = \lim_{\rho \to 0} S_\textrm{reg} = \lim_{\rho \to 0} ( S_\textrm{bulk} + S_\textrm{GH} + S_\textrm{ct} )\,,
\end{equation}
where the regularized action $S_\textrm{reg}$ is the sum of the bulk action (\ref{eq:FGPWaction}), the Gibbons-Hawking boundary term, and the counterterm action (\ref{eq:FGPWct}).
In particular, the three correlation functions are given by:
\begin{equation}
\langle O_\phi \rangle = - \lim_{\rho \to 0} \frac{\log \rho}{\rho} \frac{1}{\sqrt{\gamma}} \frac{\delta S_\textrm{reg}}{\delta \phi}\,, \quad \langle O_\psi \rangle = - \lim_{\rho \to 0} \frac{1}{\rho^{3/2}} \frac{1}{\sqrt{\gamma}} \frac{\delta S_\textrm{reg}}{\delta \psi}\,, \quad \langle T_{ij} \rangle = - \lim_{\rho \to 0} \frac{1}{\rho} \frac{2}{\sqrt{\gamma}} \frac{\delta S_\textrm{reg}}{\delta \gamma^{ij}}\,.
\end{equation}
The variation of the bulk action gives only a boundary term since the rest of the contributions are set to zero by the equations of motion.
Namely, one gets
\begin{eqnarray}
\frac{\delta S_\textrm{bulk}}{\delta \phi} & = & \frac{1}{\kappa^2} \sqrt{\gamma}\, \Big ( - \frac{2}{L} \rho \partial_\rho \phi \Big )\,, \nonumber \\
\frac{\delta S_\textrm{bulk}}{\delta \psi} & = & \frac{1}{\kappa^2} \sqrt{\gamma}\, \Big ( - \frac{2}{L} \rho \partial_\rho \psi \Big )\,, \nonumber \\
\frac{\delta S_\textrm{bulk}}{\delta \gamma^{ij}} & = & \frac{1}{2 \kappa^2} \sqrt{\gamma} \frac{\rho}{L} \big( \partial_\rho \gamma_{ij} - \gamma^{mn} \partial_\rho \gamma_{mn} \gamma_{ij} \big)\,.
\end{eqnarray}
On the other hand, the variation of the counterterm action has been already calculated during the renormalization process and it is related to the conjugate momenta of the fields:
\begin{eqnarray}
\frac{\delta S_\textrm{ct}}{\delta \phi} & = & - \pi_\phi = - \frac{1}{\kappa^2} \sqrt{\gamma} p_\phi \,, \nonumber \\
\frac{\delta S_\textrm{ct}}{\delta \psi} & = & - \pi_\psi = - \frac{1}{\kappa^2} \sqrt{\gamma} p_\psi \,, \nonumber \\
\frac{\delta S_\textrm{ct}}{\delta \gamma^{ij}} & = & - \pi_{ij} = - \frac{1}{\kappa^2} \sqrt{\gamma} \left ( Y_{ij} - \frac{1}{2} U \gamma_{ij} \right )\,.
\end{eqnarray}
After putting everything together, the following expressions are obtained:
\begin{eqnarray}
\langle O_\phi \rangle & \!\! = \!\! & - \frac{1}{\kappa^2} \lim_{\rho \to 0} \frac{\log \rho}{\rho} \left [ - \frac{2}{L}  \rho \partial_\rho \phi + \frac{2}{L} \left ( 1 + \frac{1}{\log \rho} \right ) \phi \right ] \!, \\
\langle O_\psi \rangle & \!\! = \!\! & - \frac{1}{\kappa^2} \lim_{\rho \to 0} \frac{1}{\rho^{3/2}} \left [ - \frac{2}{L} \rho \partial_\rho \psi + \frac{1}{L} \psi + \left ( \frac{1}{3 L} \left ( 1 + 3 c \right ) \psi^3 - \frac{L}{2} \left ( \square - \frac{1}{6} R \right ) \psi \right ) \log \rho \right ] \!,~~ \\
\langle T_{ij} \rangle & \!\! = \!\! & - \frac{1}{\kappa^2} \frac{2}{\rho} \left [ \frac{1}{2L} \rho \left ( \partial_\rho \gamma_{ij} - \gamma_{ij} \gamma^{mn} \partial_\rho \gamma_{mn} \right ) - Y_{ij} + \frac{1}{2} U \gamma_{ij} \right ]\!,
\label{eq:FGPWTijren}
\end{eqnarray}
with
\begin{multline}
Y_{ij} = \frac{L}{4} R_{ij} + \bigg [ \frac{L}{24} ( R_{ij} \psi^2 + 4 \nabla_i \psi \nabla_j \psi - 2 \psi \nabla_i \nabla_j \psi - ( \nabla \psi )^2 \gamma_{ij} - \psi \square \psi \gamma_{ij} ) \\
+ \frac{L^3}{96} ( 4 R R_{ij} - 12 R^{kl} R_{kilj} + \square R \gamma_{ij} + 2 \nabla_i \nabla_j R - 6 \square R_{ij} ) \bigg ] \log \rho\,,
\end{multline}
and $U$ as calculated in Section \ref{sec:FGPW}.

To determine whether the above expressions are finite, one has to use the Fefferman-Graham expansions for the metric and the scalar fields of the theory:
\begin{eqnarray}
\gamma_{ij} & \!\!\! = \!\!\! & \frac{1}{\rho} \gamma_{(0)ij} 
+ \big( \gamma_{(2)ij} + \gamma_{(2,1)ij} \log \rho \big) 
+ \rho \big( \gamma_{(4)ij} + \gamma_{(4,1)ij} \log \rho + \gamma_{(4,2)ij} \log^2 \rho \big) + \mathcal{O} ( \rho^2 ) \\
\label{psiFG}
\psi & \!\!\! = \!\!\! & \rho^{1/2} \psi_{(0)} + \rho^{3/2} \big( \psi_{(2)} + \psi_{(2,1)} \log \rho \big) + \mathcal{O} ( \rho^{5/2} ) \\[1mm]
\label{phiFG}
\phi & \!\!\! = \!\!\! & \rho \big( \phi_{(0)} + \phi_{(0,1)} \log \rho \big) + \mathcal{O} ( \rho^2 )
\end{eqnarray}
Notice that for the special case of the $\phi$-field there is a logarithmic term even in leading order in $\rho$.
(This is generally true for all fields with scaling dimension $\Delta = d / 2$.)
All the coefficients of the above expansions can be determined in terms of $\gamma_{(0) ij}$, $\gamma_{(4) ij}$, $\phi_{(0)}$, $\phi_{(0,1)}$, $\psi_{(0)}$ and $\psi_{(2)}$ using the equations of motion for the fields and the metric.
These undetermined coefficients encode information about the boundary QFT.
Namely, the leading order coefficients $\phi_{(0,1)}$ and $\psi_{(0)}$ are related to the source of the respective QFT operators, while coefficients $\phi_{(0)}$ and $\psi_{(2)}$ are related to their vev rate.
Additionally, the leading coefficient $\gamma_{(0) ij}$ in the expansion of $\gamma$ is the background metric of the boundary QFT.
Finally, although $\gamma_{(4) ij}$ is not fully determined, its trace and covariant divergence can be related to the other expansion coefficients using Einstein's equation.

The substitution of the expansion \reef{phiFG} for $\phi$ into $\langle O_\phi \rangle$ directly leads to cancellation of all of the divergences, without using the equations of motion, and the result is
\begin{equation}
\langle O_\phi \rangle = - \frac{1}{\kappa^2} \frac{2}{L} \phi_{(0)}\,.
\end{equation}

Plugging the  expansion \reef{psiFG} for $\psi$ into  $\langle O_\psi \rangle$  leads to direct cancellation of the divergent terms in leading order, i.e.~those proportional to $1 / \rho$, 
however, a  logarithmic divergence remains:
\begin{multline}
\langle O_\psi \rangle = \frac{1}{\kappa^2} \left ( \frac{2}{L} \psi_{(2)} + \frac{2}{L} \psi_{(2,1)} \right ) \\
+ \frac{1}{\kappa^2} \lim_{\rho \to 0} \left [ \frac{2}{L} \psi_{(2,1)} - \frac{1}{3 L}  ( 1 + 3 c ) \psi_{(0)}^3 + \frac{L}{2} \Big ( \square_{(0)} - \frac{1}{6} R_{(0)} \Big ) \psi_{(0)} \right ] \log \rho\,,
\end{multline}
where $R_{(0)} \equiv R[\gamma_{(0)}]$ is the Ricci scalar obtained by the metric $\gamma_{(0)}$ and
\be
\square_{(0)} \psi_{(0)} \equiv \frac{1}{\sqrt{\gamma_{(0)}}} \partial_i \Big ( \sqrt{\gamma_{(0)}} \gamma_{(0)}^{ij} \partial_j \psi_{(0)} \Big )\,.
\ee
In order to see the desired cancellations, one has to calculate the expansion coefficient $\psi_{(2,1)}$ via the equation of motion for the field $\psi$,
\begin{equation}
L^2 \square_\gamma \psi + 4 \rho^2 \partial_\rho^2 \psi + 4 \rho \partial_\rho \psi + 2 \rho^2 \partial_\rho \psi \operatorname{Tr} ( \gamma^{-1} \partial_\rho \gamma ) + 3 \psi - 2 c \psi^3 = 0\,.
\end{equation}
By the asymptotic expansions for $\psi$ and the metric, the terms proportional to $\rho^{3/2}$ give
\be
\psi_{(2,1)} = - \frac{1}{4} \left ( L^2 \square_{(0)} + \operatorname{Tr} ( \gamma_{(0)}^{-1} \gamma_{(2)} ) - 2 c \psi_{(0)}^2 \right ) \psi_{(0)}\,.
\ee
Finally, $\gamma_{(2)}$ is determined using Einstein's equation:
\begin{equation}
\mathcal{R}_{\mu \nu} [ g ] = \partial_\mu \phi \partial_\nu \phi + \partial_\mu \psi \partial_\nu \psi + \frac{1}{3 L^2} V ( \phi, \psi ) g_{\mu \nu} \,.
\end{equation}
The $ij$ component of this equation is
\begin{equation}
\begin{split}
L^2 R_{ij} [ \gamma ] =&~ 2 \rho^2 \partial_\rho^2 \gamma_{ij} + 2 \rho \partial_\rho \gamma_{ij} +  \rho^2 \operatorname{Tr} ( \gamma^{-1} \partial_\rho \gamma ) \partial_\rho \gamma_{ij} - 2 \rho^2 \gamma^{mn} \partial_\rho \gamma_{mi} \partial_\rho \gamma_{nj} \\
&- \frac{1}{2} \rho^2 \operatorname{Tr} ( \gamma^{-1} \partial_\rho \gamma^{-1} \partial_\rho \gamma ) \gamma_{ij} + \rho^2 \operatorname{Tr} ( \gamma^{-1} \partial_\rho^2 \gamma ) \gamma_{ij} + \rho \operatorname{Tr} ( \gamma^{-1} \partial_\rho \gamma ) \gamma_{ij} \\
&+ L^2 \partial_i \phi \partial_j \phi +  L^2 \partial_i \psi \partial_j \psi  + 2 \rho^2 ( \partial_\rho \psi )^2 \gamma_{ij} + 2 \rho^2 ( \partial_\rho \psi )^2 \gamma_{ij} + \frac{2}{L^2} V ( \phi, \psi ) \gamma_{ij}\,.
\end{split}
\end{equation}
Expanding it and keeping terms up to $\mathcal{O} ( 1 ) $ one finds
\be
\gamma_{(2) ij} = - \frac{L^2}{2} \bigg ( R_{(0) ij} - \frac{1}{6} R_{(0)} \gamma_{(0) ij} \bigg ) - \frac{1}{6} \psi_{(0)}^2 \gamma_{(0) ij}.
\ee
Now using these results for $\psi_{(2,1)}$ and $\gamma_{(2)}$ in $\langle O_\psi \rangle$ exactly cancels the logarithmic term and gives the following finite result for the one-point function:
\begin{equation}
\langle O_\psi \rangle = \frac{1}{\kappa^2} \left [ \frac{2}{L} \psi_{(2)} - \frac{L}{2} \Big ( \square_{(0)} - \frac{1}{6} R_{(0)} \Big ) \psi_{(0)} + \frac{1}{3L} ( 1 + 3 c ) \psi_{(0)}^3 \right ] \,.
\end{equation}

A similar approach leads to the renormalized one-point function for the energy-momentum tensor.
A direct substitution of the asymptotic expansions in equation~(\ref{eq:FGPWTijren}) leads to the cancellation of the leading $\mathcal{O} ( \rho^{-2} )$ divergences.
However, the remaining divergences can be canceled only after solving Einstein's equation for $\gamma_{(4,1)}$ and $\gamma_{(4,2)}$.
Terms proportional to $\rho \log \rho$ give
\begin{equation}
\gamma_{(4,2) ij} = - \frac{1}{6} \phi_{(0,1)}^2 \gamma_{(0) ij} \,,
\end{equation}
while  terms proportional to $\rho$ give 
\begin{equation}
\begin{aligned}
\gamma_{(4,1) ij} = & \frac{L^4}{8} \left ( R_{(0)}^{kl} R_{(0) ikjl} - \frac{1}{3} R_{(0)} R_{(0) ij} \right ) - \frac{L^4}{32} \left ( R_{(0)}^{kl} R_{(0) kl} - \frac{1}{3} R_{(0)}^2 \right ) \gamma_{(0) ij} \\
& + \frac{L^4}{16} \left ( \square_{(0)} R_{(0) ij} - \frac{1}{3} \nabla_i \nabla_j R_{(0)} - \frac{1}{6} \square_{(0)} R_{(0)} \gamma_{(0) ij} \right ) \\
& + \frac{L^2}{4} \psi_{(0)} \left ( \frac{1}{3} \nabla_i \nabla_j + \frac{1}{6} \gamma_{(0) ij} \square_{(0)} - \frac{1}{6} R_{(0) ij} \right ) \psi_{(0)} \\
& - \frac{L^2}{6} \left ( \nabla_i \psi_{(0)} \nabla_j \psi_{(0)} - \frac{1}{4} \gamma_{(0)}^{kl} \nabla_k \psi_{(0)} \nabla_l \psi_{(0)} \gamma_{(0) ij} \right ) \\
& - \frac{1}{24} \left ( 1 + 3 c \right ) \psi_{(0)}^4 \gamma_{(0) ij} - \frac{1}{3} \phi_{(0)} \phi_{(0,1)} \gamma_{(0) ij}\,.
\end{aligned}
\end{equation}
Then, the renormalized energy momentum tensor will be given by:
\begin{equation}
\begin{aligned}
\langle T_{ij} \rangle = & - \frac{2}{L} \gamma_{(4) ij} - \frac{1}{L} \left (  \frac{1}{3} \phi_{(0)}^2 - \phi_{(0)} \phi_{(0,1)} + \frac{2}{3} \phi_{(0,1)}^2 - \frac{1}{72} \left ( 1 - 3 c \right ) \psi_{(0)}^4 + \psi_{(0)} \psi_{(2)} \right ) \gamma_{(0) ij} \\
& + \frac{L}{8} \left ( \gamma_{(0)}^{kl} \nabla_k \psi_{(0)} \nabla_l \psi_{(0)} + \psi_{(0)} \left ( \square_{(0)} - \frac{1}{9} R_{(0)} \right ) \psi_{(0)} \right ) \gamma_{(0) ij} \\
& - \frac{L}{4} \psi_{(0)} \left ( \nabla_i \nabla_j - \frac{1}{2} R_{(0) ij} \right ) \psi_{(0)}  + \frac{L^3}{32} \left ( R_{(0) kl} R_{(0)}^{kl} + \frac{1}{9} R_{(0)}^2 + \square_{(0)} R_{(0)} \right ) \gamma_{(0) ij} \\
& + \frac{L^3}{4} \left ( \tensor{R}{_{(0) i}^k} R_{(0) kj} - \frac{3}{2} R_{(0)}^{kl} R_{(0) ikjl} + \frac{1}{4} \nabla_i \nabla_j R_{(0)} - \frac{3}{4} \square_{(0)} R_{(0) ij} \right )\,.
\end{aligned}
\end{equation}
The trace of the stress-tensor one-point function gives a much simpler expression, since the trace $\operatorname{Tr} ( \gamma_{(0)}^{-1} \gamma_{(4)} )$ can be obtained from the $\rho \rho$ component of Einstein's equation, which gives
\begin{equation}
\rho^2 \operatorname{Tr} ( \gamma^{-1} \partial_\rho \gamma \gamma^{-1} \partial_\rho \gamma ) - 2 \rho^2 \operatorname{Tr} ( \gamma^{-1} \partial_\rho^2 \gamma ) - 2 \rho \operatorname{Tr} ( \gamma^{-1} \partial_\rho \gamma ) = ( 2 \rho \partial_\rho \phi )^2 + ( 2 \rho \partial_\rho \psi )^2 + \frac{L^2}{3} V ( \phi, \psi )\,.
\end{equation}
Keeping only terms of order $\mathcal{O} ( \rho^2 )$ in this yields
\begin{equation}
\begin{aligned}
\operatorname{Tr} ( \gamma_{(0)}^{-1} \gamma_{(4)} ) = & \frac{L^4}{16} \left ( R_{(0) ij} R_{(0)}^{ij} - \frac{2}{9} R_{(0)}^2 \right ) - \frac{L^2}{8} \psi_{(0)} \left ( \square_{(0)} - \frac{5}{18} R_{(0)} \right ) \psi_{(0)} \\
& - \frac{1}{3} ( 2 \phi_{(0)}^2 + \phi_{(0,1)}^2 ) + \frac{1}{9} \left ( 1 + \frac{3}{2} c \right ) \psi_{(0)}^4 - \psi_{(0)} \psi_{(2)}\,.
\end{aligned}
\end{equation}
After plugging in the above result the trace anomaly becomes
\begin{multline}
\langle \tensor{T}{^i_i} \rangle = \frac{1}{L} \left ( 4 \phi_{(0)} \phi_{(0,1)} - 2 \phi_{(0,1)}^2 - \frac{1}{6} \left ( 1 + 3 c \right ) \psi_{(0)}^4 - 2 \psi_{(0)} \psi_{(2)} \right ) \\
+ \frac{L}{2} \left ( \psi_{(0)} \square_{(0)} \psi_{(0)} + \gamma_{(0)}^{ij} \partial_i \psi_{(0)} \partial_j \psi_{(0)} \right ) - \frac{L^3}{8} \left ( R_{(0) ij} R_{(0)}^{ij} - \frac{1}{3} R_{(0)}^2 \right )\,.
\end{multline}

It must be mentioned that the above results for the one-point functions are true only up to contributions from finite counterterms in the action.

\end{document}